\documentclass[aps,prb,twocolumn, groupedaddress, showpacs]{revtex4}
\usepackage{amsmath}
\usepackage{amssymb}
\usepackage{graphicx}
\usepackage{color}
\usepackage{tikz}
\usepackage{pgffor}
\usepackage{verbatim}
\newcommand{\<}{\langle}
\renewcommand{\>}{\rangle}
\newcommand{\be}{\begin{equation} }
\newcommand{\ee}{\end{equation} }
\newcommand{\ba}{\begin{eqnarray} }
\newcommand{\ea}{\end{eqnarray} }

\newcommand{\down}{\downarrow}

\newcommand{\bpm}{\begin{pmatrix}}
\newcommand{\epm}{\end{pmatrix}}
\newcommand{\bmm}{\begin{matrix}}
\newcommand{\emm}{\end{matrix}}
\newcommand{\up}{\uparrow}

\newtheorem{thm}{Theorem}

\begin{document}

\title{Weak symmetry breaking in two dimensional topological insulators}

\author{Chenjie Wang}
\altaffiliation{{\it Current address: James Franck Institute, Department of Physics,
University of Chicago, Chicago, Illinois 60637, USA.}}
\affiliation{Condensed Matter Theory Center, Department of Physics, University of Maryland, College Park, Maryland 20742, USA}

\author{Michael Levin}
\altaffiliation{{\it Current address: James Franck Institute, Department of Physics,
University of Chicago, Chicago, Illinois 60637, USA.}}
\affiliation{Condensed Matter Theory Center, Department of Physics, University of Maryland, College Park, Maryland 20742, USA}

\date{\today}

\begin{abstract}
We show that there exist two dimensional (2D) time reversal invariant fractionalized insulators with the property that both their boundary with the vacuum and their boundary with a topological insulator can be fully gapped without breaking time reversal or charge conservation symmetry. This result leads us to an apparent paradox: we consider a geometry in which a disk-like region made up of a topological insulator is surrounded by an annular strip of a fractionalized insulator, which is in turn surrounded by the vacuum. If we gap both boundaries of the strip, we naively obtain an example of a gapped interface between a topological insulator and the vacuum that does not break any symmetries -- an impossibility. The resolution of this paradox is that this system spontaneously breaks time reversal symmetry in an unusual way, which we call \emph{weak symmetry breaking}. In particular, we find that the only order parameters that are sensitive to the symmetry breaking are nonlocal operators that describe quasiparticle tunneling processes between the two edges of the strip; expectation values of local order parameters vanish exponentially in the limit of a wide strip. Also, we find that the symmetry breaking in our system comes with a ground state degeneracy, but this ground state degeneracy is topologically protected, rather than symmetry protected. We show that this kind of symmetry breaking can also occur at the edge of 2D fractional topological insulators.
\end{abstract}


\maketitle

\section{Introduction}


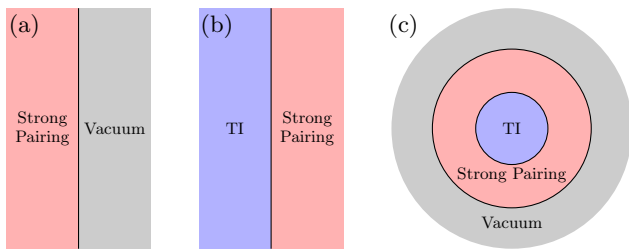
\begin{figure}[b]
\centering
\begin{tikzpicture}[scale=1.6]
\fill [red!30](0,0) rectangle +(-0.6,2);
\fill [black!20] (0,0) rectangle +(0.6,2);
\draw (0,0)--(0,2);
\node at (-0.3, 1)[align=center, scale=0.7]{Strong \\Pairing};
\node at (0.3, 1)[align=center, scale=0.7]{Vacuum};
\node at (-0.45, 1.85){(a)};

\def\px {1.6}

\fill [blue!30](0+\px,0) rectangle +(-0.6,2);
\fill [red!30] (0+\px,0) rectangle +(0.6,2);
\draw (0+\px,0)--(0+\px,2);
\node at (0.3+\px, 1)[align=center, scale=0.7]{Strong \\Pairing};
\node at (-0.3+\px, 1)[align=center, scale=0.7]{TI};
\node at (-0.45+\px, 1.85){(b)};

\def\ppx {3.6}
\def\ppy {1}

\fill [black!20](0+\ppx,0+\ppy)  circle(1);
\draw [fill=red!30](0+\ppx,0+\ppy) circle(0.66);
\draw [fill=blue!30](0+\ppx,0+\ppy) circle(0.3);
\node at(0+\ppx,0+\ppy)[scale=0.7]{TI};
\node at(0+\ppx, -0.39+\ppy)[scale=0.7]{Strong Pairing};
\node at(0+\ppx, -0.78+\ppy)[scale=0.7]{Vacuum};
\node at (-0.9+\ppx, .85+\ppy){(c)};
\end{tikzpicture}
\caption{(a) Boundary between a strong pairing insulator (red) and the vacuum (gray). (b) Boundary between a topological insulator (blue) and a strong pairing insulator. (c) A geometry that contains both boundaries, in which the strong pairing strip can be considered as a  broadened boundary between the topological insulator and the vacuum.}
\label{fig1}
\end{figure}

One of the most important distinctions between two dimensional topological insulators and two dimensional trivial insulators is that the interface between a topological insulator and the vacuum carries robust gapless edge modes \cite{kane05a, kane05b, bernevig06, hasan10} while no such modes are present for a trivial insulator. The edge modes of a topological insulator are protected by the fundamental symmetries of time reversal and charge conservation \cite{kane05b, xu06, wu06}. That is, it is impossible to fully gap out the edge of a topological insulator without breaking at least one of these symmetries, either explicitly or spontaneously.

Inspired by topological insulators, searches for such ``symmetry protected'' edge modes have been carried out in other systems. One extension is to systems with other symmetries\cite{schnyder08, kitaev09, chen13, lu12} beyond time reversal and charge conservation symmetry. Another important extension is to systems with intrinsic topological order\cite{levin09, levin12, neupert11, lu13}, that is systems that support bulk excitations with fractional statistics. In both cases, one of the central issues is the stability of the edge modes. Because of recent intensive studies\cite{fidkowski10, ryu12, yao12, qi13, gu13, levin09, levin12,neupert11, lu12, lu13}, much of the stability issue has been understood in a systematic way. Of particular relevance to this work are Refs.~\onlinecite{levin09, levin12} , which analyzed the edge stability of 2D time reversal invariant Abelian insulators. Here, by ``Abelian'' insulators, we mean insulators whose fundamental bulk excitations obey Abelian fractional statistics. Ref.~\onlinecite{levin09, levin12} found that time reversal invariant Abelian insulators can be divided into fractional topological insulators and fractional trivial insulators according to the edge stability, similar to the usual ``unfractionalized'' insulators.


In this paper, we follow the work of Ref.~\onlinecite{levin09} and \onlinecite{levin12}, and continue to study the edge stability of two dimensional time reversal invariant Abelian insulators and interesting phenomena resulting from the stability study. While Ref.~\onlinecite{levin09} and \onlinecite{levin12} focused on edge modes that live at the interface between an insulator and the vacuum, the same approach can be used to study boundaries between two different insulators. As we show below, the analysis of these boundaries yields a surprising result: there exist time reversal invariant fractionalized insulators with the property that \emph{both} their boundary with the vacuum (Fig.~\ref{fig1}a) and their boundary with a conventional topological insulator (Fig.~\ref{fig1}b) can be fully gapped by appropriate edge interactions -- without breaking time reversal or charge conservation symmetry. An example of such a fractionalized insulator is the ``strong-pairing insulator'' which we define in Sec.~\ref{strongpairing}. To understand why this result is surprising, note that such a scenario is impossible for time reversal invariant band insulators: in such systems, a boundary can be gapped if and only if the two neighboring phases have the same $Z_2$ invariant. Hence, there are no band insulators with the property that both their boundary with the vacuum and their boundary with a topological insulator can be gapped without breaking any symmetries.

The purpose of this paper is to derive this result and resolve a paradox associated with it. The paradox occurs when we consider a geometry in which a disk-like region made up of a conventional topological insulator is surrounded by an annular strip of the strong-pairing insulator, which is in turn surrounded by the vacuum (Fig.~\ref{fig1}c). This system has an energy gap everywhere except at the topological insulator/strong pairing insulator boundary, and at the strong pairing insulator/vacuum boundary; these interfaces may or may not support gapless edge modes, depending on what interactions are present nearby. Let us suppose that the annular strip is much wider than the microscopic correlation length, so that these two boundaries decouple from another and can be treated independently. Then, according to the above result, we can introduce edge interactions which will fully gap out both the topological insulator/strong pairing insulator boundary and the strong pairing insulator/vacuum boundary without breaking time reversal or charge conservation symmetry. This scenario leads us to an apparent contradiction: thinking of the annular strip as a wide edge, we have seemingly constructed a fully gapped interface between a topological insulator and the vacuum that does not break any symmetries!

The resolution of this paradox is subtle. We find that the strip spontaneously breaks time reversal symmetry, but this symmetry breaking has an unusual character. In particular, we find that the symmetry breaking in the strip cannot be detected by local order parameters. More precisely, the expectation value of any local order parameter is extremely small -- exponentially small in the width of the strip. The only order parameters that are sensitive to the symmetry breaking are \emph{nonlocal} string-like operators that describe tunneling processes between the two edges of the strip. Closely related to this, we find that, unlike traditional symmetry breaking, there are multiple degenerate ground states and this ground state degeneracy cannot be split even if time reversal symmetry is explicitly broken. Indeed, we find that the ground state degeneracy is a topological degeneracy that originates from the topological order in the annular strip, rather than a symmetry breaking degeneracy. We call this unusual kind of symmetry breaking {\it weak symmetry breaking}\cite{footnote1} and we show that there is a crossover between weak symmetry breaking and conventional symmetry breaking as the width of the annular strip is reduced: in the narrow-strip limit, the broken time reversal symmetry becomes detectable by local order parameters and the topological degeneracy becomes a symmetry breaking degeneracy.

The paper is organized as follows. In Sec.~\ref{example}, we construct the strong pairing insulator and show that its boundary with the vacuum and its boundary with a topological insulator can be fully gapped without breaking any symmetries. We then derive the resulting paradox and discuss its resolution in terms of weak symmetry breaking. Then we move on to analyze the weak symmetry breaking. We discuss the existence of nonlocal order parameters in Sec.~\ref{orderparameter}, discuss ground state degeneracy as well as finite size corrections in Sec.~\ref{degeneracy}, and discuss the exponential suppression of local order parameters in Sec.~\ref{localorderparam}. The discussion is generalized to fractional topological insulators in Sec.~\ref{general}, with a general nonlocal order parameter found in Sec.~\ref{general3}. We conclude in Sec.~\ref{conclusion}. In appendix \ref{degAppend}, we obtain general formulas for the ground state degeneracy of gapped edges in various geometries, and in appendix \ref{append_a} we prove a theorem for Abelian spin Hall insulators.

\section{Derivation of paradox}
\label{example}

In this section, we derive the paradox through a detailed study of the boundaries in Fig.~\ref{fig1}a, Fig.~\ref{fig1}b and Fig.~\ref{fig1}c respectively.

\subsection{The strong pairing insulator}
\label{strongpairing}
First, we construct a toy model for the strong pairing insulator. To this end, we recall that a toy model for a conventional 2D topological insulator can be obtained by considering a state in which spin-up and spin-down electrons form decoupled $\nu = 1$ integer quantum Hall states with opposite chiralities. The wave function for this state is given by:
\begin{align}
\psi(\{z^\up_i, & z^\down_i\}) =\prod_{i < j} (z^\up_i - z^\up_j) \prod_{i < j} (\bar{z}^\down_i - \bar{z}^\down_j) \nonumber \\ &
\times \exp\left[-\sum_{i}\left(|z_i^\uparrow|^2+| z_i^\downarrow|^2\right)/4l_B^2\right],
\end{align}
where $z^\up_i$, $z^\down_i$ denote the coordinates of the spin-up and spin-down electrons, $l_B$ denotes the magnetic length, and $\bar z_i^\downarrow$ means the complex conjugate of $z_i^\downarrow$. (We leave the anti-symmetrization between the two spin species implicit, as is standard for multi-component/multi-layer quantum Hall wave functions). This state is called a ``quantum spin Hall'' insulator since it exhibits a nonvanishing spin-Hall conductivity: $\sigma_{sH} = \nu = 1$ (in units of $e/2\pi$).

By generalizing this construction, one can easily obtain toy models for time reversal invariant \emph{fractionalized} insulators\cite{bernevig06, levin09, neupert11}. Indeed, to construct such states, we simply imagine that the spin-up and spin-down electrons form {\it fractional} quantum Hall states with opposite chiralities. These states are called ``fractional quantum spin Hall'' insulators since they have a fractional spin-Hall conductivity $\sigma_{sH}$.

The ``strong pairing'' insulator is a particular fractional quantum spin Hall state in which the spin-up and spin-down electrons form $\nu = 1/2$ strong pairing FQH states with opposite chiralities. Here, the $\nu = 1/2$ strong pairing FQH state is an Abelian fractional quantum Hall state in which spin-polarized electrons first bind together to form charge $2e$ Cooper pairs, and then the pairs form a $k = 8$ bosonic Laughlin state. The wave function for the strong pairing insulator state is given by
\begin{align}
\psi (\{w^\up_i, & w^\down_i\}) = \prod_{i < j} (w^\up_i - w^\up_j)^8 \prod_{i < j} (\bar{w}^\down_i - \bar{w}^\down_j)^8 \nonumber \\ &
\times \exp\left[-\sum_{i}\left(|w_i^\uparrow|^2+| w_i^\downarrow|^2\right)/4l_B^2\right],
\end{align}
where  $w^\up_i$, $w^\down_i$ denote the coordinates of the charge $2e$, spin polarized Cooper pairs formed out of the spin-up and spin-down electrons.

\subsection{Boundary between strong pairing insulator and vacuum}
\label{sp-vacuum}

\begin{figure}[b]
\centering
\begin{tikzpicture}[scale=1.5]
\fill [red!30](0,0) rectangle +(-0.6,2);
\fill [black!20] (0,0) rectangle +(0.6,2);
\draw [-stealth] (-0.08,0.05)--(-0.08,1.95);
\draw [-stealth] (-0.14,1.95)--(-0.14,0.05);
\node at (-0.45, 1.85){(a)};

\def\px {1.6}

\fill [blue!30](0+\px,0) rectangle +(-0.6,2);
\fill [red!30] (0+\px,0) rectangle +(0.6,2);
\draw [-stealth] (-0.08+\px,0.05)--(-0.08+\px,1.95);
\draw [-stealth] (-0.14+\px,1.95)--(-0.14+\px,0.05);

\draw [-stealth] (0.08+\px,0.05)--(0.08+\px,1.95);
\draw [-stealth] (0.14+\px,1.95)--(0.14+\px,0.05);

\node at (-0.45+\px, 1.85){(b)};

\def\ppx {3.2}
\def\ppy {1}

\fill [blue!30](0+\ppx,0) rectangle +(-0.6,2);
\fill [red!30] (0+\ppx,0) rectangle +(0.6,2);
\fill [black!20] (0.6+\ppx,0) rectangle +(0.6,2);

\draw [-stealth] (-0.08+\ppx,0.05)--(-0.08+\ppx,1.95);
\draw [-stealth] (-0.14+\ppx,1.95)--(-0.14+\ppx,0.05);

\draw [-stealth] (0.08+\ppx,0.05)--(0.08+\ppx,1.95);
\draw [-stealth] (0.14+\ppx,1.95)--(0.14+\ppx,0.05);

\draw [-stealth] (0.52+\ppx,0.05)--(0.52+\ppx,1.95);
\draw [-stealth] (0.46+\ppx,1.95)--(0.46+\ppx,0.05);

\node at (-0.45+\ppx, .85+\ppy){(c)};

\node at (-0.11, -0.2){$\phi_2\ \phi_1$};
\node at (0+\px, -0.2){$\phi_6\ \phi_5\ \phi_4\ \phi_3$};
\node at (0.3+\ppx, -0.2){$\phi_6\ \phi_5\ \phi_4\ \phi_3\ \phi_2\ \phi_1$};

\end{tikzpicture}
\caption{Schematics of boundary modes. (a) The boundary between the strong pairing insulator and the vacuum has two modes, $\phi_1,\phi_2$. (b) The boundary between the strong pairing insulator and the topological insulator has four modes, $\phi_3,\phi_4, \phi_5, \phi_6$. (c) The combination of the two boundaries has six modes, $\phi_1,\phi_2, \phi_3,\phi_4, \phi_5, \phi_6$. (Labeling starts from the right.) }
\label{fig2}
\end{figure}
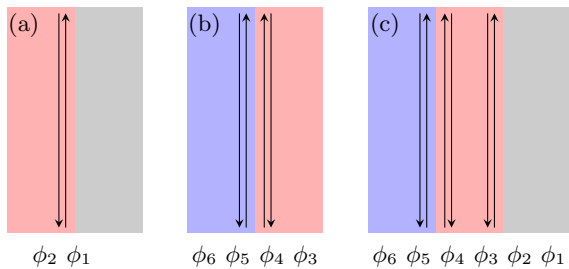

In this section we analyze the boundary between the strong pairing insulator and the vacuum (Fig.~\ref{fig1}a). We show that this boundary can be fully gapped without violating time reversal and charge conservation symmetry. In other words, we show that the strong pairing insulator does not have symmetry-protected edge modes. (There is a simple criterion for determining the stability of this boundary and its generalization (see Sec.~\ref{review}), however here we will give a detailed analysis.)

To begin, let us recall the edge theory of the strong pairing FQH state. Since the strong pairing FQH state is a $k=8$ Laughlin state made out of charge $2e$ Cooper pairs, its edge can be modeled by a single chiral boson mode $\phi$ with a Lagrangian
\begin{equation}
L= \frac{8}{4\pi}\left[\partial_t\phi \partial_x\phi -v(\partial_x\phi)^2\right].
\end{equation}
The edge of the strong pairing insulator can be modeled as two copies of the strong pairing FQH edge:
\begin{eqnarray}
L_a &=& \frac{8}{4\pi}\left[\partial_t\phi_1\partial_x\phi_1-v(\partial_x\phi_1)^2\right] \nonumber \\
&+& \frac{8}{4\pi}\left[-\partial_t\phi_2\partial_x\phi_2-v(\partial_x\phi_2)^2\right] . \label{sp-edge}
\end{eqnarray}
Here $\phi_1$ and $\phi_2$ describe the spin-up and spin-down edge modes respectively (Fig.~\ref{fig2}a). In this normalization convention, the creation operator for spin-up, charge $2e$ Cooper pairs is of the form $b_\up^\dagger \sim e^{8i\phi_1}$ while the creation operator for spin-down Cooper pairs is $b_\down^\dagger \sim e^{-8i \phi_2}$.

We will adopt the convention that $\phi_1$ and $\phi_2$ transform under time reversal symmetry as
\begin{align}
\phi_1 \rightarrow \phi_2 \ , \ \phi_2 \rightarrow \phi_1.
\label{trtrans}
\end{align}
This transformation law guarantees that the Cooper pair creation operators transform correctly under time reversal, namely $b_\up^\dagger \rightarrow b_\down^\dagger$, and $b_\down^\dagger \rightarrow b_\up^\dagger$.

The above Lagrangian can be written in a compact form using the so called $K$-matrix formalism\cite{wen, wen92, wen95}:
\begin{equation}
L=\frac{1}{4\pi} (\partial_t\Phi^T \mathcal K \partial_x \Phi - \partial_x\Phi^T \mathcal V
\partial_x\Phi ),
\label{k-lagrangian}
\end{equation}
with $\mathcal K$ a symmetric nonsingular integer matrix, $\mathcal V$ a positive-definite velocity matrix, and $\Phi$ a multi-component field. In our case of the strong pairing insulator, the Lagrangian $L_a$ in (\ref{sp-edge}) is written in the $K$-matrix formalism with
\begin{align}
\mathcal K = \mathcal K_a = \bpm  8 & 0  \\
                                  0 & -8  \epm \  ,
\ \mathcal V = \mathcal V_a = \bpm  8v & 0 \\
                                    0 & 8v  \epm,
\end{align}
and
\begin{align}
\Phi = \Phi_a = \bpm \phi_1 \\ \phi_2 \epm.
\end{align}
In this notation, a general product of spin-up and spin-down Cooper pair creation and annihilation operators can be written in the form $e^{i \Lambda^T \mathcal{K}_a \Phi_a}$ where $\Lambda$ is a two component integer vector.

Our aim is to show that the two edge modes in Eq.~(\ref{sp-edge}) can be gapped out by suitable perturbations, without breaking time reversal or charge conservation symmetry. We will now argue that the following perturbation does the job:
\begin{align}
U_{a} = U(x)\cos(\Lambda_1^T \mathcal{K}_a \Phi_a - \alpha(x)), \quad
\Lambda_1=\left(
\begin{matrix}
1\\
-1
\end{matrix}
\right).
\label{pert-a}
\end{align}
To see this, notice that this term describes a spin-flip process in which a spin-up/spin-down Cooper pair is destroyed and a spin-down/spin-up Cooper pair is created. Thus this term is charge conserving. Furthermore, it is time reversal invariant according to the transformation (\ref{trtrans}). To see that it gaps out the edge, we make a change of variables to $\theta = 8\phi_1 + 8\phi_2$, $\varphi = \frac{1}{2}(\phi_1 - \phi_2)$, which transforms the Lagrangian (\ref{sp-edge}) to a sine-Gordon model:
\begin{align}
L_{a} - U_{a}& = \frac{1}{2\pi} \partial_x \varphi \partial_t \theta - \frac{v}{4\pi} \left[\frac{(\partial_x \theta)^2}{16} + 16(\partial_x \varphi)^2 \right] \nonumber \\
 & - U(x)\cos(\theta - \alpha(x))
\end{align}
It is then clear that for large $U$, the field $\theta$ will become locked to the minimum of the cosine potential and the edge will be gapped.

In principle, we should also check that this perturbation does not break any symmetry \emph{spontaneously}. We will verify this in Sec.~\ref{breaking} and will postpone the discussion of spontaneous breaking until then.

\subsection{Boundary between strong pairing insulator and topological insulator}
\label{sp-ti}
In this section we show that the boundary between the strong pairing insulator and a conventional topological insulator (Fig.~\ref{fig1}b) also does not have symmetry-protected edge modes. (Again, there is a simple criterion for determining the stability of this boundary (see Sec.~\ref{review}), but we give a more detailed analysis here.)

The boundary between these two states can be modeled by four edge modes $\phi_3, \phi_4, \phi_5, \phi_6$ (Fig.~\ref{fig2}b), with a Lagrangian of the form
\begin{eqnarray}
L_{b} &=& \frac{8}{4\pi}\left[-\partial_t\phi_3\partial_x\phi_3-v'(\partial_x\phi_3)^2\right] \nonumber \\
&+& \frac{8}{4\pi}\left[\partial_t\phi_4\partial_x\phi_4-v'(\partial_x\phi_4)^2\right] \nonumber \\
&+& \frac{1}{4\pi}\left[\partial_t\phi_5\partial_x\phi_5-v''(\partial_x\phi_5)^2\right] \nonumber \\
&+& \frac{1}{4\pi}\left[-\partial_t\phi_6\partial_x\phi_6-v''(\partial_x\phi_6)^2\right]. \label{sp-ti-edge}
\end{eqnarray}
Here $\phi_3, \phi_4$ are the spin-up and spin-down edge modes of the strong pairing insulator and $\phi_5, \phi_6$ are the spin-up and spin-down edge modes of the topological insulator. Our normalization convention is such that the creation operators for spin-up and spin-down electrons in the topological insulator edge are given by $\psi_\up^\dagger \sim e^{i\phi_5}$, $\psi_\down^\dagger \sim e^{-i\phi_6}$. Similarly, the Cooper pair creation operators on the strong pairing edge are given by $b_\up^\dagger \sim e^{-8i\phi_3}$, $b_\down^\dagger \sim e^{8i\phi_4}$.

Alternatively, the Lagrangian $L_b$ can be written in the $K$-matrix formalism (\ref{k-lagrangian}) with
\begin{align}
\mathcal K_{b} = \bpm -8 & 0 & 0 & 0 \\
                   		    0 & 8 & 0 & 0 \\
                   		    0 & 0 & 1 & 0 \\
                   		    0 & 0 & 0 & -1 \epm,  \, \
\mathcal V_{b} = \bpm  8v' & 0 & 0 & 0 \\
                   		    0 & 8v' & 0 & 0 \\
                   		    0 & 0 & v'' & 0 \\
                   		    0 & 0 & 0 & v'' \epm,
\end{align}
and
\begin{align}
\Phi_{b} = \bpm \phi_3 \\ \phi_4 \\ \phi_5 \\ \phi_6 \epm.
\end{align}

We use the same convention for the time reversal transformation law of the strong pairing edge modes $\phi_3, \phi_4$ as in Eq. (\ref{trtrans}):
\begin{align}
\phi_3 \rightarrow \phi_4 \ , \ \phi_4 \rightarrow \phi_3. \label{trtrans1}
\end{align}
On the other hand, we assume that the topological insulator edge modes $\phi_5, \phi_6$ transform as:
\begin{align}
\phi_5 \rightarrow \phi_6 \ , \ \phi_6 \rightarrow \phi_5-\pi.
\label{trtrans2}
\end{align}
The extra $\pi$ in this transformation law is necessary because the electron creation operators should transform as $\psi_\up^\dagger \rightarrow \psi_\down^\dagger$, and $\psi_\down^\dagger \rightarrow -\psi_\up^\dagger$.

Our goal is to show that the gapless edge modes (\ref{sp-ti-edge}) can be gapped out by an appropriate local perturbation without breaking time reversal or charge conservation symmetry, explicitly or spontaneously. As before, we will accomplish this gapping by adding backscattering terms similar to (\ref{pert-a}). Since there are $4$ gapless edge modes, we need to add two such backscattering terms. We will now argue that the following two scattering terms do the job:
\begin{align}
U_{b} & = U(x)[\cos(\Lambda_2^T\mathcal K_b \Phi_b -\alpha(x)) \nonumber \\ &\quad \quad - \cos(\Lambda_3^T\mathcal K_b \Phi_b -\alpha(x))], \nonumber \\
\Lambda_2 & = \left(
\begin{matrix}
0 \\
1 \\
1 \\
-3
\end{matrix}
\right), \
\Lambda_3=\left(\begin{matrix}
-1 \\
0  \\
3  \\
-1
\end{matrix}
\right).
\label{pert-b}
\end{align}
These terms have all the required properties. First of all, these perturbations are local, i.e., they are composed out of products of electron creation/annihilation operators acting near some point $x$ in space. Indeed, in the $K$-matrix formalism, local operators -- i.e. those composed out of electron creation/annihilation operators -- are of the form $e^{i\Lambda^T\mathcal K \Phi}$ with $\Lambda$ an integer vector. (On the other hand, quasiparticle creation and annihilation operators, which are generally non-local, take the form $e^{i l^T \Phi}$ where $l$ is an integer vector). Second, we can see that both terms are neutral and hence preserve the $U(1)$ charge symmetry. This can be seen from the physical picture of the perturbations. For example, the $\Lambda_2$ term corresponds to a process in which one spin-down Cooper pair is created in the strong pairing edge while one spin-up electron is created and three spin-down electrons are annihilated in the topological insulator edge, or vice versa. So this term is clearly charge conserving. Similarly, we can see that the $\Lambda_3$ term is charge conserving. Third, these terms are time reversal symmetric according to the time reversal transformation (\ref{trtrans1}) and (\ref{trtrans2}).

Now that we have established that the perturbations (\ref{pert-b}) have the required symmetry properties, we have two questions
to answer: (i) whether the perturbations will gap the edge and (ii) if so, whether any symmetry is spontaneously broken as a
consequence of this gapping. To answer the first question, we use the null vector criterion of Ref.~\onlinecite{haldane95}.
According to this criterion, a perturbation of the form (\ref{pert-b}) will gap out the edge for large $U$ if and only if
\begin{equation}
\Lambda_2^T\mathcal K_b \Lambda_2 = \Lambda_3^T\mathcal K_b \Lambda_3 = \Lambda_2^T \mathcal K_b \Lambda_3 =0.
\label{nullcond}
\end{equation}
The origin of the null vector criterion is that it guarantees that one can make a linear change of variables from $\Phi_b$ to
$\Phi'_b$ such that the Lagrangian for $\Phi'_b$ will be equivalent to two decoupled sine-Gordon models. It is then clear that
if $U$ is sufficiently large, the two combinations $\Lambda_2^T \mathcal K_b \Phi_b$, $\Lambda_3^T \mathcal K_b \Phi_b$ will
become locked to the minima of the cosine potential and the edge will be gapped. (See appendix \ref{degAppend} for a derivation).

One may easily check that $\Lambda_2, \Lambda_3$ satisfy the null vector condition (\ref{nullcond}). We conclude that the
boundary can indeed be gapped by (\ref{pert-b}). As for question (ii) regarding spontaneous symmetry breaking, we will discuss
this issue in the next subsection.

\subsection{Absence of spontaneous symmetry breaking on both boundaries}
\label{breaking}

To determine whether the two gapped boundaries studied above spontaneously break any symmetries, we use a general criterion
introduced by Ref.~\onlinecite{levin12}. This criterion, which we will call the ``primitivity criterion'' can be stated as
follows: Consider a general $K$-matrix edge theory (\ref{k-lagrangian}) with a $2N\times 2N$ $K$-matrix and perturbations
\begin{equation}U_1\cos(\Lambda_1^T\mathcal K \Phi-\alpha_1), \ \dots,\ U_N\cos(\Lambda_N^T\mathcal K \Phi-\alpha_N).
\label{perturbgen}
\end{equation}
Suppose that the $\Lambda_i$ obey the null vector condition $\Lambda_i^T \mathcal K \Lambda_j=0$, so that these perturbations can
gap the edge. The primitivity criterion states that the resulting gapped edge will not break any symmetry spontaneously if
$\{\Lambda_i\}$ are ``primitive.'' Here, we say that an integer vector $\Lambda$ is primitive if it cannot be written as
an integer multiple of another integer vector -- i.e. $\Lambda \neq k \Lambda'$ for any integer $k$ and integer vector
$\Lambda'$; similarly, a vector set $\{\Lambda_i\}$ is primitive if all linear combinations $\sum_{i}a_i\Lambda_i$ are primitive
for any integers $a_1,\dots,a_N$ with no common divisor. A simple way to check the primitivity of $\{\Lambda_i\}$
is to see if the set of $N\times N$  minors of the matrix $\mathcal  M = (\Lambda_1, \dots \Lambda_N)$ are relatively prime:
Ref.~\onlinecite{levin12} showed that the set of $2N$ dimensional vectors $\{\Lambda_1,\dots, \Lambda_N\}$ is primitive if and
only if the $N\times N$ minors of $\mathcal M$ have no common divisor.

The basic intuition behind the primitivity condition is that it checks for the existence of a local order parameter. We can see
this with a simple example: suppose $\Lambda = k \Lambda'$ so that $\Lambda$ is not primitive. Then when the cosine term
$\cos(\Lambda^T \mathcal K \Phi)$ locks the value of $\Lambda^T \mathcal K \Phi$ to its minima, it will also freeze the value of
$(\Lambda')^T \mathcal K \Phi$. We can then construct an operator of the form $e^{i (\Lambda')^T \mathcal K \Phi -i\alpha}$ that
will have a nonzero expectation value in the ground state. Furthermore, since $\Lambda'$ is a \emph{fraction} of $\Lambda$, this
operator can transform nontrivially under symmetries that leave $\cos(\Lambda_i^T \mathcal K \Phi)$ invariant, implying that a
symmetry is broken spontaneously. In this way, we see that if the set of $\{\Lambda_i\}$ is not primitive, then a local order
parameter can be constructed and spontaneous symmetry breaking is possible. On the other hand, if $\{\Lambda_i\}$ is primitive
then no order parameter can be constructed (or at least no order parameter of the form $e^{i \Lambda^T \mathcal K \Phi}$), and
thus spontaneous symmetry breaking is not possible.

A rigorous derivation of the primitivity condition can be obtained using the ground state degeneracy formula (\ref{gsd}) from section
\ref{degeneracy1} (see also Eq.~(\ref{diskdeg}) in appendix \ref{degAppend}). That formula applies to Abelian states in a disk geometry with an edge that has been gapped by perturbations of the form (\ref{perturbgen}). It states that the ground state degeneracy of such a system is equal to the
greatest common divisor of the set of $N \times N$ minors of the matrix $\mathcal M = (\Lambda_1,\dots, \Lambda_N)$. Given this
formula, it is easy to establish the primitivity condition: we can see that the primitivity condition guarantees that the ground
state of our system is \emph{non-degenerate}. It then follows that no symmetry is broken spontaneously, since any spontaneously
broken symmetry would necessarily be accompanied by either a ground state degeneracy or gapless excitations, and our system is
gapped by assumption.

We now apply the primitivity condition to the two gapped boundaries studied above. We can see that $\{\Lambda_2, \Lambda_3\}$ is
primitive by direct calculation of the $2 \times 2$ minors of the matrix $\mathcal M = (\Lambda_2, \Lambda_3)$. Also, it is
clear from inspection that $\Lambda_1$ is primitive. We conclude that neither of the two boundaries studied in the previous two
sections spontaneously break any symmetries.

\subsection{The paradox}
\label{paradox}

We have shown that \emph{both} the boundary between the strong pairing insulator and the vacuum and the boundary between the strong pairing insulator and the topological insulator can be gapped without breaking any symmetry. We will show that this result leads us to an apparent paradox.

As discussed in the introduction, the paradox occurs when we consider a geometry in which a disk-like region filled with a topological insulator is surrounded by an annular strip filled with a strong pairing insulator which is in turn surrounded by the vacuum (Fig.~\ref{fig1}c). Let us call the outer boundary between the strong pairing insulator and the vacuum {\it boundary $a$}, and call the inner boundary between the strong paring insulator and the topological insulator {\it boundary $b$}. Clearly, boundary $a$ can be modelled in the same way as the boundary in Fig.~\ref{fig1}a, with two edge modes $\phi_1, \phi_2$ and the edge Lagrangian $L_a[\phi_1, \phi_2]$ in (\ref{sp-edge}). Likewise, boundary $b$ can be modelled in the same way as the boundary in Fig.~\ref{fig1}b, with four edge modes $\phi_3, \phi_4, \phi_5, \phi_6$ and the Lagrangian $L_b[\phi_3, \phi_4, \phi_5, \phi_6]$ in (\ref{sp-ti-edge}) (See Fig.~\ref{fig2}c). Assuming that the annular strip is much wider than the microscopic correlation length, we can neglect coupling between the two boundaries. The total Lagrangian for the annular strip is then given by
\begin{equation}
L = L_a[\phi_1, \phi_2] + L_b[\phi_3, \phi_4, \phi_5, \phi_6].
\end{equation}
Note that both $\phi_1$ and $\phi_3$ are edge modes of the spin-up component of the strong pairing insulator, but they are located at opposite sides of the strip, so they have opposite chiralities. It is the same for $\phi_2$ and $\phi_4$. Such opposite chirality is reflected in our normalization conventions in (\ref{sp-edge}) and (\ref{sp-ti-edge}) which differ by a minus sign.

In Sec.~\ref{sp-vacuum}, we constructed a perturbation $U_a$ which gaps out the two modes $\phi_1$, $\phi_2$ at the strong-pairing insulator/vacuum boundary. In Sec.~\ref{sp-ti}, we constructed a perturbation $U_b$ which gaps out the four modes $\phi_3$, $\phi_4$, $\phi_5$, $\phi_6$ at the strong-pairing insulator/topological insulator boundary. Furthermore, we showed that neither of these gappings lead to the breaking of time reversal or charge conservation symmetry -- either explicitly or spontaneously.

Let us now imagine adding the perturbations $U_a$ and $U_b$ to boundary $a$ and $b$ respectively in the geometry Fig.~\ref{fig1}c. With exactly the same analysis in Sec.~\ref{sp-vacuum} and Sec.~\ref{sp-ti}, we know that both boundaries $a$ and $b$ are gapped and no symmetry breaking occurs. If we think of the annulus as a wide interface between a topological insulator and the vacuum, then we seem to have found a way to gap the edge of the topological insulator without breaking any of the fundamental symmetries -- an impossibility. In what follows we will show how to resolve this apparent paradox.

\section{Nonlocal Order Parameter}
\label{orderparameter}

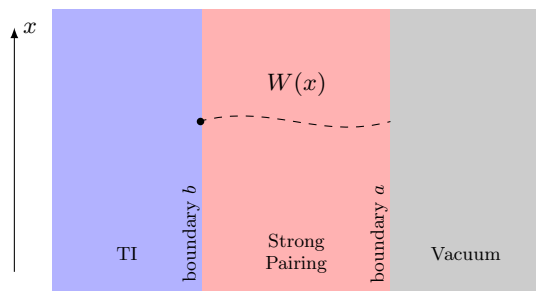
\begin{figure}[b]
\begin{tikzpicture}[scale=2.5]

\fill [blue!30](0,0) rectangle +(0.8,1.5);
\fill [red!30] (0.8,0) rectangle +(1,1.5);
\fill [black!20] (1.8,0) rectangle +(0.8,1.5);

\draw [-latex](-0.2,0.1)--(-0.2,1.4) node [anchor=west]{$x$};
\draw [dashed](0.78, 0.9) to [out=15, in=195](1.8, 0.9);
\fill (0.79,0.9) circle(0.02);

\node at (0.4, 0.2)[scale=0.8]{TI};
\node at (1.3, 0.2)[align=center, scale=0.8]{Strong \\ Pairing};
\node at (2.2, 0.2)[scale=0.8]{Vacuum};
\node at (1.3, 1.1){$W(x)$};

\node at (0.74, 0.3)[rotate=90, scale=0.8]{boundary $b$};
\node at (1.74, 0.3)[rotate=90, scale=0.8]{boundary $a$};
\end{tikzpicture}
\caption{Physical picture of the nonlocal order parameter $W(x)$, Eq.~(\ref{string-example}): it moves a quasiparticle $(2,2)^T$ between the two boundaries, and at the same time it flips an electron spin at boundary $b$ (black dot).}\label{fig3}
\end{figure}

Now we resolve the paradox associated with the geometry in Fig.~\ref{fig1}c. The resolution involves an unusual kind of breaking of time reversal symmetry, which we call {\it weak symmetry breaking}. In this and the next two sections, we will prove that the annular strip undergoes a spontaneous breaking of time reversal symmetry, and we will characterize the ``weakness'' of such breaking. In this section, we obtain a nonlocal order parameter as the first evidence of the weak breaking of time reversal symmetry.

\subsection{Existence of order parameter}
\label{orderparameter12}

In Sec. \ref{sp-vacuum} and \ref{sp-ti}, we analyzed boundary $a$ and boundary $b$ separately and we showed that no spontaneous symmetry breaking occurs when either of these boundaries is gapped. Now let us consider the geometry of
Fig.~\ref{fig1}c where boundaries $a$ and $b$ are connected by an annular strip. If the strip is very wide,
the two boundaries must decouple from one another, and thus it is clear that there cannot be symmetry
breaking at either of the two gapped boundaries. In other words, we expect that there are no local order parameters
acting near either of the two boundaries that acquire a nonzero ground state expectation value. (For a more
detailed argument against the existence of local order parameters, see Sec.~\ref{localorderparam}).

Given that there are no local order parameters, it is tempting to conclude that the system in Fig.~\ref{fig1}c
does not break any symmetries. However this conclusion is incorrect. The problem is that checking for local order
parameters is not enough to determine the existence of spontaneous symmetry breaking: we also need to
include \emph{nonlocal} operators such as quasiparticle tunneling operators between the two boundaries. Including such
nonlocal observables is the key to resolving the paradox. In fact, it is natural to include these nonlocal
observables: The paradox occurs when we think the annular strip as a wide edge of the topological insulator,
so we should include all operators acting within the strip.

We now demonstrate that there exists a nonlocal order parameter of time reversal symmetry breaking, i.e., a nonlocal
operator that is odd under time reversal and that acquires a nonvanishing ground state expectation value.
This order parameter is defined by
\begin{equation}
W(x)=\cos(2\phi_1+2\phi_2+2\phi_3+2\phi_4+\phi_5+\phi_6) \label{string-example}
\end{equation}
where the position variable $x$ is written explicitly to emphasize that $W$ can live anywhere along the
boundaries. To understand the physical meaning of $W$, recall that in the $K$-matrix formalism, the operator
$e^{il^T\Phi}$ creates a quasiparticle excitation $l$ at the boundary. Thus, $e^{i2\phi_1+i2\phi_2}$ creates
a quasiparticle $(2,2)^T$ at boundary $a$ while $e^{i2\phi_3+i2\phi_4}$ annihilates a quasiparticle $(2,2)^T$ at
boundary $b$. Also, the operator $e^{i\phi_5+i\phi_6}$ corresponds to an electron spin flip. Putting these all
together, we see that $W$ describes a physical process in which a quasiparticle $l=(2,2)^T$ tunnels from boundary
$a$ to boundary $b$, or vice versa, and at the same time an electron flips its spin at boundary $b$ (Fig.~\ref{fig3}).

Let us verify that $W(x)$ is indeed an order parameter of time reversal symmetry. First, we note that
$W(x)\rightarrow -W(x)$ under time reversal transformation (\ref{trtrans}), (\ref{trtrans1}) and (\ref{trtrans2}).
Second, to see that $W$ acquires a ground state expectation value when boundaries $a$ and $b$ are gapped, note that
one may write $W$ as
\begin{equation}
W= \cos\left[\frac{1}{4}\left(\Lambda_1^T\mathcal K_a\Phi_a+\Lambda_2^T\mathcal K_b\Phi_b+\Lambda_3^T\mathcal K_b\Phi_b\right)\right],
\end{equation}
following the notations from equations (\ref{pert-a}) and (\ref{pert-b}). In the large $U$ limit, the
fields $\Lambda_1^T\mathcal K_a\Phi_a$, $\Lambda_2^T\mathcal K_b\Phi_b$ and $\Lambda_3^T\mathcal K_b\Phi_b$ are
locked at the minima of the cosine potentials, therefore $W$ is locked to a classical value. In general, the
expectation value of $W$ is nonvanishing, because the phase $\alpha(x)$ in (\ref{pert-a}) or (\ref{pert-b}) is
arbitrary.  Hence, $W(x)$ is indeed an order parameter characterizing the breaking of time reversal symmetry.

\subsection{Nonlocality of order parameter}
\label{scale}


\begin{figure}
\begin{tikzpicture}[scale=2.3]
\draw [black!70,fill=black!70, rotate=175] (0.5,0) circle(0.03);
\draw [black!70, fill=black!70, rotate=170] (1,0) circle(0.03);

\draw (0,0)  circle(1);
\draw (0,0) circle(0.5);
\node at(0.9,0.1)[scale=1]{$  S^{\rm II}_l$};
\node at(0.65, -0.6)[scale=1]{$  S^{\rm I}_l$};
\draw [dashed,  latex-](0.5,0) to [out=30, in =-150](1,0);
\draw [dashed, ] (0,0)circle (0.75);
\draw [-latex] (0.05, -0.75)--(0.1, -0.747);
\draw [fill=black] (0.5,0) circle(0.02);
\draw [fill=black] (1,0) circle(0.02);

\node at (-1.1, 0.3)[scale=1]{$U_a$};
\node at (-0.35, 0.15)[scale=1]{$U_b$};

\draw [stealth-stealth, double, rotate=210] (0.5,0)--(1,0);
\draw [rotate=210] (0.8,0) node[anchor=south, rotate=30]{$D$};

\end{tikzpicture}
\caption{Local operators ($U_a$, $U_b$) and string operators ($ S^{\rm I}_l$, $ S^{\rm II}_l$). There are two types of non-contractible string operators, type-I and type-II, associated with moving quasiparticles along the two dashed lines respectively. The width of the annular strip is $D$ while the average length of the boundaries is $L$.}
\label{fig4}
\end{figure}
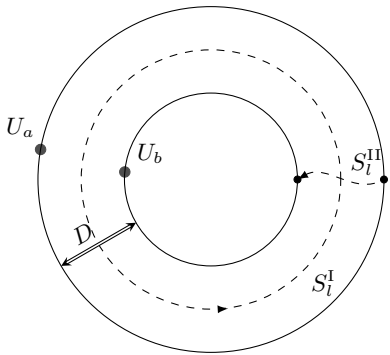

The nonlocality of the order parameter $W$ is clear from the physical picture behind it: it involves quasiparticle tunneling across the annular strip. However, $W$ in (\ref{string-example}) is expressed in the edge theory, in which the bulk degrees of freedom are projected out. The edge theory cannot tell the distance between the boundaries and the width of the annular strip does not appear explicitly. This makes it ambiguous to talk about the nonlocality of $W$. Therefore, it is necessary to clarify this conceptual ambiguity. In particular, we discuss below how the width of the strip enters our edge theory analysis.

To begin, we compare the three characteristic lengths in our problem: the microscopic correlation length $\xi$, the width of the strip $D$, and the average length of the boundaries $L$. In the toy models in Sec.~\ref{strongpairing}, the microscopic scale $\xi$ is the magnetic length $l_B$. In our analysis, we always assume $\xi \ll L$. The ratio of $D$ to $\xi$ gives two limits of our problem: the wide-strip limit $D\gg \xi$, and the narrow-strip limit $D\sim \xi$. Up to now, we have assumed the wide-strip limit.

Now we clarify the concepts of {\it local} and {\it nonlocal} operators. Local operators are those acting on a region whose size is comparable to $\xi$, for examples, $U_a$ in (\ref{pert-a}) which acts somewhere at boundary $a$ and $U_b$ in (\ref{pert-b}) which acts somewhere at boundary $b$ (see schematics in Fig.~\ref{fig4}). They are composed out of the electron creation/annihilation operators located within the region where these local operators are defined. Nonlocal operators act on a region whose size is much larger than the microscopic scale. The string-like operators $S_l^{\rm I}$ and $S_l^{\rm II}$ sketched in Fig.~\ref{fig4}, called type-I and type-II string operators, are our main examples of nonlocal operators -- they will be useful in the study of the ground state degeneracy in Sec.~\ref{degeneracy}. They describe a quasiparticle $l$ tunneling along the dashed lines in Fig.~\ref{fig4}. $S_l^{\rm I}$ acts on the scale $L$ and $S_l^{\rm II}$ acts on the scale $D$. Obviously $S_l^{\rm I}$ is nonlocal. Whether $S_l^{\rm II}$ is local or nonlocal depends on what limits we take: it is local in the narrow-strip limit ($D\sim \xi$) and it is nonlocal in the wide-strip limit ($D \gg \xi$).

With the above preparation, we are ready to discuss the locality/nonlocality of the order parameter. As mentioned above, the operator $W$ is defined in the edge Hilbert space where the notion of the width of the strip fades away, so it is ambiguous to talk about the locality/nonlocality of $W$. However, since physically $W$ describes quasiparticle tunneling across the annular strip, there must exist a $S_l^{\rm II}$-like string operator $\tilde W$ in the full Hilbert space whose projection in the edge Hilbert space is $W$. $\tilde W$ and $W$ are the {\it same} physical quantity (the order parameter) but expressed in different representations. Introducing the $\tilde W$ representation of the order parameter makes it easy to understand its locality/nonlocality: Similarly to $S_l^{\rm II}$, the order parameter is nonlocal in the wide-strip limit, but it becomes a local quantity in the narrow-strip limit. This clarifies what we mean by nonlocality of the order parameter in the wide-strip limit. In what follows, we will abuse notation and not distinguish $W$ from $\tilde W$ as long as no confusion occurs. We will use the phrases ``$W$ is local'' or ``$W$ is nonlocal'' with no further explanation.

A natural question is then: How is the locality/nonlocality of the order parameter manifested in the edge theory? One manifestation is that the operator $W$ enters the edge Hamiltonian with only exponentially small coefficients in the wide-strip limit.  Imagine we project $\tilde W$ to $W$ in the edge Hilbert space. In the wide-strip limit, $\tilde W$ does not enter the microscopic Hamiltonian. Furthermore, the bulk of the strip is gapped. Thus, $W$ can only be generated from a quasiparticle tunneling process and a rough estimate of the coefficient of $W$ in the effective edge Hamiltonian is of the order of $\exp(-\alpha D )$, with $\alpha$ a positive number. In the limit $D/\xi\rightarrow \infty$, the coefficient vanishes and $W$ is forbidden from entering the edge Hamiltonian. We see the width of the annular strip $D$ enters the edge theory in this rather implicit way.

Finally, we make a comment about the narrow-strip limit. In that limit, the order parameter $W$ becomes local, and therefore the symmetry breaking becomes locally detectable. The symmetry breaking becomes ``strong''. Hence, reducing the width of the strip provides a natural crossover from the weak breaking to the usual breaking of time reversal symmetry. Such a crossover will be discussed in more detail in Sec.~\ref{crossover} from the view of ground state degeneracy. Strictly speaking, whenever we say weak symmetry breaking, we stay in or close to the wide-strip limit.

\section{Ground State Degeneracy}
\label{degeneracy}
It is well known that systems with a spontaneously broken discrete symmetry necessarily exhibit a ground state degeneracy. Here we show that the weak symmetry breaking in our system (Fig. \ref{fig1}c) also leads to a ground state degeneracy. However, the ground state degeneracy differs from the usual symmetry breaking case in several ways. First, we find a degeneracy of $4$ instead of the two-fold degeneracy one expects for time reversal symmetry breaking. Second, we find that the degeneracy cannot be split by any local perturbation of the Hamiltonian, including perturbations that break time reversal symmetry. Thus, the ground state degeneracy should be regarded as a \emph{topological} degeneracy, not a symmetry-breaking degeneracy. Both of these unusual features disappear in the narrow-strip limit: in that case, we find a crossover to a two-fold degeneracy that is protected by time reversal symmetry.

\subsection{Computing the degeneracy}
\label{degeneracy1}

In this section we calculate the ground state degeneracy of our system (Fig.~\ref{fig1}c) in the wide-strip limit. Our calculation is based on a formula derived in appendix \ref{degAppend}. This formula gives the ground state degeneracy of a general system with the geometry of Fig.~\ref{fig1}c, in which both the disk and the annulus contain Abelian states and all boundaries are gapped. Before explaining this result, we first discuss a simpler formula that gives the ground state degeneracy of a general Abelian state defined in a \emph{disk} geometry with a gapped edge.

The formula for the ground state degeneracy in a disk geometry is as follows. Consider an Abelian state described by a $2N \times 2N$ $K$-matrix
$\mathcal{K}$. Suppose that this state is defined in a disk, with an edge gapped by backscattering terms
\begin{equation}
U_1\cos(\Lambda_1^T \mathcal K \Phi), \dots, U_N\cos(\Lambda_N^T\mathcal K \Phi), \label{gapcondition}
\end{equation}
where $\{\Lambda_i\}$ are $2N$-dimensional linearly-independent integer vectors that satisfy the null vector condition $\Lambda_i^T \mathcal K \Lambda_j =0$. Then, the ground state degeneracy (GSD) is given by
\begin{equation}
{\rm GSD_{disk}}= {\rm gcd}(\text{$N\times N$ minors of } \mathcal M), \label{gsd}
\end{equation}
where $\rm gcd$ stands for the ``greatest common divisor'' and $\mathcal M$ is the $2N\times N$ matrix whose columns are the vectors $\Lambda_i$
\begin{equation}
\mathcal M=  (\Lambda_1, \dots,\Lambda_N).
\end{equation}
Note that equation (\ref{gsd}) gives the \emph{total} ground state degeneracy of our system, not just the topologically protected degeneracy\cite{jwang13, wen89}, which is presumably trivial for a disk-like system.

A detailed derivation of the formula (\ref{gsd}) is given in appendix \ref{degAppend}. Here we provide an intuitive, but less rigorous explanation. First, consider the structure of the edge Hilbert space before we add the backscattering terms (\ref{gapcondition}). In this case, the following operators commute with the Hamiltonian:
\begin{equation}
P_I = \frac{1}{2\pi} \int_0^L dx \partial_x \Phi_I.
\end{equation}
These operators have integer eigenvalues $p_I$, so every eigenstate can be labeled by a $2N$-component integer vector $p$. Eigenstates with the same values of $p$ are said to belong to the same ``sector.'' Each sector contains a lowest energy state together with an infinite tower of excited states associated with phonon excitations.

Let us now add the backscattering terms (\ref{gapcondition}). These terms have two effects. First, the combinations $\Lambda_i^T \mathcal K \Phi$ become locked to the minima of the cosine potential at low energies. Therefore, $\Lambda_i^T \mathcal K \partial_x \Phi = 0$, so the only sectors that have low energy states are those with
\begin{equation}
\Lambda_i^T \mathcal K p = 0,
\label{gammalat}
\end{equation}
for all $\Lambda_i$. Let us denote the set of all integer vectors $p$ that satisfy  equation (\ref{gammalat}) by $\Gamma$. The second effect is that the $p$'s are no longer good quantum numbers since the cosine terms couple states in sector $p$ with states in sector $p + \Lambda_i$. To obtain good quantum numbers, we need to consider $p \text{ (mod $A$)}$ where $A$ is the $N$ dimensional lattice generated by $\Lambda_1,...,\Lambda_N$. Note that $A\subseteq \Gamma$ by the null vector criterion.

To complete the derivation, we note that it is natural to expect that there is one degenerate ground state for each of the low energy sectors. In other words, there is one ground state for each element of $\Gamma/A$. It is then not hard to show that the number of elements of $\Gamma/A$ is precisely equal to the expression in equation (\ref{gsd}), thus proving the claim.

Having ``warmed up'' with the disk case, we now introduce the formula for the ground state degeneracy for general Abelian states defined in the geometry of Fig. \ref{fig1}c. Suppose that the annular strip is described by an Abelian state with a $2N\times 2N$ $K$-matrix $\mathcal K_s$, while the disk is described by an Abelian state with a $2M\times 2M$ $K$-matrix $\mathcal K_d$. Suppose that the outer boundary $a$ and the inner boundary $b$ are gapped by backscattering terms corresponding to $\{\Lambda_{a,1},...,\Lambda_{a,N}\}$ and $\{\Lambda_{b,1},...,\Lambda_{b,N+M}\}$ where the $\Lambda_{a,i}$ are $2N$-dimensional, and the $\Lambda_{b,i}$ are $(2M+2N)$-dimensional integer vectors. Then the ground state degeneracy is given by:
\begin{equation}
{\rm GSD_{Fig.~\ref{fig1}c}}= {\rm gcd}(\text{$(2N+M)\times (2N+M)$ minors of } \mathcal M') \label{gsdfig1c}
\end{equation}
where $\mathcal M'$ is a $(4N + 2M) \times (2N + M)$ matrix. The first $N$ columns of $\mathcal{M'}$ are
\begin{align}
\left(\begin{matrix}
\mathcal K_s\Lambda_{a,i} \\
\Lambda_{a,i} \\
0_{2M}
\end{matrix}
\right), \ \ i = 1,..., N,
\end{align}
where $0_{2M}$ is the $2M$-dimensional zero vector while the last $M+N$ columns are
\begin{align}
\left(\begin{matrix}
0_{2N} \\
\Lambda_{b,i}
\end{matrix}
\right), \ \ i = 1,..., M+N.
\end{align}

The physics behind equation (\ref{gsdfig1c}) is similar to the disk case and we refer the reader to appendix \ref{degAppend} for
its derivation. Here, we will simply apply this formula to compute the degeneracy in our system. In our case, we have
\begin{eqnarray}
\Lambda_{a,1}^T &=& (1,-1), \nonumber \\
\Lambda_{b,1}^T &=& (0,1,1,-3), \nonumber \\
\Lambda_{b,2}^T &=& (-1,0,3,-1),
\end{eqnarray}
and $K_s$ = $\text{diag}(8,-8)$. Therefore, the matrix $\mathcal M'$ is given by
\begin{equation}
\mathcal M' = \left(
\begin{matrix}
8 & 0 & 0 \\
8 & 0 & 0 \\
1 & 0 & -1\\
-1 & 1 & 0\\
0 & 1 & 3\\
0 & -3 & -1
\end{matrix}
\right).
\end{equation}
We conclude that the ground state degeneracy for our system is
\begin{align}
{\rm GSD_{Fig.~\ref{fig1}c}}= {\rm gcd}(\text{$3\times3$ minors of $\mathcal M'$}) = 4.
\end{align}

\subsection{Topological origin of the degeneracy}

In the previous section we showed that the system drawn in Fig.~\ref{fig1}c exhibits a four-fold ground state degeneracy. The next
question is to determine the robustness of this degeneracy. Here we show that the degeneracy cannot be split by any local perturbation
of the Hamiltonian, including perturbations that break time reversal symmetry. We conclude that the ground state degeneracy should be thought of
as a topological degeneracy -- analogous to the degeneracy of a topologically ordered state on a torus, rather than a symmetry-breaking degeneracy.

Our argument is similar to previous approaches for establishing topological degeneracy \cite{wen89, jwang13}: our strategy is to construct string-like operators
that (1) satisfy a nontrivial commutation algebra, and (2) map ground states to ground states. We then use these two properties to prove that
the ground state degeneracy is robust.

We make use of two types of string-like operators. The first type
of operator, which we call $S_l^{\rm I}$, acts along a non-contractible loop in the annular strip (i.e. the dashed circle in
Fig.~\ref{fig4}). Physically this operator describes a three step process in which one first creates a quasiparticle/quasihole pair
$l$/$\bar l$, then winds $l$ around the central disk, and finally brings $l$ back to $\bar l$ and annihilates the pair. The second type
of string-like operator, which we call $S_l^{\rm II}$, acts along a path connecting the inner and outer boundaries of the strip. This
operator describes a process in which one creates a pair $l/\bar l$, then takes $\bar l$ to boundary $a$ and $l$ to boundary $b$, and
finally annihilates them separately at the two boundaries.

A key property of these two types of string operators is that they don't commute with each other in general. Instead they satisfy
the algebra
\begin{align}
S_l^{\rm I}S_{l'}^{\rm II} &= e^{i\theta_{ll'}} S_{l'}^{\rm II}S_{l}^{\rm I}
\label{stralgebra}
\end{align}
where $\theta_{ll'}$ is the mutual statistical phase between quasiparticle $l$ and $l'$. To see this, note that
$(S_{l}^{\rm I})^{-1}(S_{l'}^{\rm II})^{-1}S_l^{\rm I}S_{l'}^{\rm II}$ describes a process in which quasiparticle
$l$ is braided around quasiparticle $l'$, followed by a process in which quasiparticle $l$ is braided in the opposite
direction around an empty disk. The difference between the phases acquired during these two processes is exactly the
mutual statistics $e^{i\theta_{ll'}}$.

We now construct the above algebra (\ref{stralgebra}) more explicitly.
First, we need to parameterize the different string operators in our system. We begin with the type-I string
operators. Recall that in the strong pairing insulator, the quasiparticle excitations are labeled by two
component integer vectors $l$. Therefore, all the $S_l^{\rm I}$ can be built out of two
generating string operators, $S_1^{\rm I}$ and $S_2^{\rm I}$, corresponding to $l^T = (1,0)$ and $l^T = (0,1)$ respectively.

Next consider the type-II string operators, $S_l^{\rm II}$. An important point is that $S_l^{\rm II}$ can only be
constructed if the quasiparticle $l$ and quasihole $\bar{l}$ can be locally annihilated at the boundaries of the annular strip; in general this local annihilation is only possible for a subset of quasiparticles\cite{levin13}.
In the present case, we will argue that the subset of quasiparticles that can be annihilated includes $l^T = (2,2)$ and its multiples. To see this, consider the following operator defined within the edge theory:
\begin{equation}
S = \exp[-i(2\phi_1+2\phi_2+2\phi_3+2\phi_4)].
\end{equation}
The operator $S$ is a string-like operator that describes a process in which a quasiparticle $(2, 2)^T$ tunnels from boundary $a$ to boundary $b$.
If we apply $S$ to one of the ground states of our system, it will create an excited state with a quasiparticle $(2,2)^T$ on boundary $b$ and
a corresponding quasihole on boundary $a$. In order to construct a type-II string operator out of $S$, we need to construct local operators acting on the two boundaries that can annihilate these excitations and return the system to its ground state. We claim that the follow operator
does the job:
\begin{equation}
V = \exp[-i (\phi_5 + \phi_6)].
\end{equation}
Clearly $V$ is a local operator that describes a single electron flipping its spin on boundary $b$. Furthermore, consider the product $V \cdot S$.
We can see that $V \cdot S$ looks like the nonlocal order parameter $W$ in Eq.~(\ref{string-example}) -- in fact $W=(V\cdot S + S^\dagger \cdot V^\dagger)/2$. Therefore,
just like $W$, $V \cdot S$ acquires a ground state expectation value when the edges are gapped by the backscattering terms (\ref{pert-a})
and (\ref{pert-b}). The fact that $V \cdot S$ acquires an expectation value tells us that if we apply $V \cdot S$ to a ground state of our
system, the resulting state has a nonvanishing overlap with the ground state subspace. In other words, the operator $V$ gives some amplitude
for annihilating the quasiparticle $(2,2)^T$ at the boundary. We conclude that $(2,2)^T$ can indeed be annihilated at the boundary.
In addition, we have an explicit expression for the corresponding string operator, which we denote by $S^{\rm II}_0$:
\begin{eqnarray}
S^{\rm II}_0 &\sim& V \cdot S \label{s2def} \\
&=& \exp[-i(2\phi_1+2\phi_2+2\phi_3+2\phi_4 + \phi_5 + \phi_6)].\nonumber
\end{eqnarray}
We will find this expression useful in Sec.~\ref{parameterization}.

We have constructed three fundamental string operators:
$S_1^{\rm I}$, $S_2^{\rm I}$, $S_0^{\rm II}$. To find their commutation algebra (\ref{stralgebra}),
we recall that according to the $K$-matrix formalism, the mutual statistics between two excitations $l, l'$ of the strong pairing state is
given by
\begin{equation}
\theta_{ll'} = 2\pi l^T \bpm 8 & 0 \\ 0 & -8 \epm ^{-1} l'.
\end{equation}
Substituting $l^T = (1,0), (0,1),(2,2)$, we derive the algebra
\begin{align}
 S_1^{\rm I} S_0^{\rm II} &= e^{i\pi/2}  S_0^{\rm II}  S_1^{\rm I},\nonumber\\
 S_2^{\rm I} S_0^{\rm II} &= e^{-i\pi/2} S_0^{\rm II}  S_2^{\rm I}.
\label{stralgebra2}
\end{align}

With the above algebra in hand, we are now ready to complete the argument. The key point is that each of the string
operators $S_1^{\rm I}$, $S_2^{\rm I}$, $S_0^{\rm II}$ map ground states to ground states, so the ground state subspace must
provide a representation of the algebra (\ref{stralgebra}). Given that the smallest representation of this algebra
is four dimensional, we immediately deduce that the ground state subspace must have degeneracy of at least $4$. Furthermore,
the same reasoning holds even if we add some arbitrary local perturbation to the Hamiltonian: our argument only depends on the
quasiparticle statistics in the bulk and the basic structure of edge, both of which are expected to be stable to arbitrary
local perturbations of the Hamiltonian (as long as the perturbations are not so large that they close the bulk or edge gap).
We conclude that the system has a minimal four-fold degeneracy that cannot be split by any local perturbation in the wide-strip limit. In other words,
the four-fold ground state degeneracy of our system is \emph{topologically protected}.

\subsection{Parameterization of degenerate ground states}
\label{parameterization}
We will find it useful to construct explicit matrix representations of the above string operators within the
four dimensional ground state subspace. We begin with $S_0^{\rm II}$. From the algebra (\ref{stralgebra2}), we can see
that the eigenvalues of this operator are necessarily of the form $\{\lambda, \lambda e^{i\pi /2}, \lambda e^{i\pi}, \lambda e^{3i\pi /2}\}$.
In fact, since we are free to rescale $S_0^{\rm II}$ by a complex scalar, we can assume without loss of generality that the
eigenvalues of $S_0^{\rm II}$ are of the form $e^{in\pi/2}$ with $n = 0,1,2,3$. Working in this eigenbasis, we have
\begin{equation}
S_0^{\rm II}  =\left( \begin{array} {cccc}
1 & 0 & 0 & 0 \\
0 & i & 0 & 0 \\
0 & 0 & -1 & 0 \\
0 & 0 & 0 & -i
\end{array} \right).
\label{mrep}
\end{equation}
Similarly, from the algebra (\ref{stralgebra2}) we deduce that
\begin{equation}
S_1^{\rm I} = (S_2^{\rm II})^{-1} = \bpm 0 & 0 & 0 & 1 \\
					1 & 0 & 0 & 0 \\
					0 & 1 & 0 & 0 \\
					0 & 0 & 1 & 0 \epm
\end{equation}
(after an appropriate rescaling). Labeling these basis states by $|n\>$, $n = 0,1,2,3$,
we can equivalently write these equations as
\begin{align}
S_0^{\rm II}|n\rangle &= e^{i\pi n/2}|n\rangle,\nonumber\\
S_1^{\rm I}|n\rangle &= |n-1\rangle, \nonumber\\
S_2^{\rm I}|n\rangle &=  |n+1\rangle, \label{string-gss}
\end{align}
where the addition and subtraction in the last two equations is performed modulo $4$.
We can also obtain the matrix representation
of the non-local order parameter $W$ using the fact that $W \sim (S_0^{\rm II} + (S_0^{\rm II})^{\dagger})$:
\begin{equation}
W |n\rangle \sim \cos(n\pi/2)|n\rangle.
\end{equation}

At this point it is worth mentioning an important property of the string operators, namely,
their action within the ground state subspace does not change if we deform the path along
which they act. More precisely, if we let $S_{0 \gamma}^{\rm II}$ denote
the string operator corresponding to a particular path $\gamma$, then
\begin{equation}
S_{0 \gamma}^{\rm II}|n\> = S_{0 \gamma'}^{\rm II}|n\>,
\label{pathind}
\end{equation}
for any two paths $\gamma, \gamma'$ connecting the two boundaries of the strip. The other string operators,
$S_1^{\rm I}, S_2^{\rm I}$ exhibit a similar path independence. To derive (\ref{pathind}), note that
$S_{0\gamma}^{\rm II}$ and $S_{0\gamma'}^{\rm II}$ obey the same algebra (\ref{stralgebra2}), so
the operator $S_{0\gamma}^{\rm II} \cdot (S_{0\gamma'}^{\rm II})^{-1}$ commutes with $S_1^{\rm I}$. At
the same time, it is clear that $S_{0\gamma}^{\rm II} \cdot (S_{0\gamma'}^{\rm II})^{-1}$ commutes with
$S_{0\gamma}^{\rm II}$ (at least if $\gamma, \gamma'$ do not intersect). These two facts imply that
$S_{0\gamma}^{\rm II}\cdot (S_{0\gamma'}^{\rm II})^{-1} = C I$ where $I$ is the $4 \times 4$ identity
and $C$ is some complex scalar. Hence, as long as we rescale the two operators appropriately so that $C=1$,
they will obey (\ref{pathind}).

To complete the picture, we now construct an explicit matrix representation for the time reversal transformation $\mathcal T$. According to
the time reversal transformations (\ref{trtrans}), (\ref{trtrans1}), (\ref{trtrans2}), and the expression (\ref{s2def}), we have
$\mathcal T^{-1} S_0^{\rm II} \mathcal T = - S_0^{\rm II\dag}$. It follows that
\begin{equation}
\mathcal T |n\rangle \sim |n+2\rangle.
\end{equation}
Choosing an appropriate phase convention for $|n\>$, we conclude that the time reversal operator can be represented within the ground state
subspace as
\begin{align}
\mathcal T =  T  K \ , \
 T  =\left( \begin{array} {cccc}
0 & 0 & 1 & 0 \\
0 & 0 & 0 & 1 \\
1 & 0 & 0 & 0 \\
0 & 1 & 0 & 0
\end{array} \right).
\label{trep}
\end{align}
where $ K$ is the anti-linear complex conjugation operator.

\subsection{Crossover to narrow-strip limit}
\label{crossover}

The above four-fold topological ground state degeneracy is obtained in the wide-strip limit. We now study the degeneracy in the narrow-strip
limit. We find that as the width of the strip is reduced, the degeneracy undergoes a crossover to a two-fold degeneracy which is protected by
time reversal symmetry. Thus, in the narrow-strip limit the ground state degeneracy behaves exactly as we would expect for a system with
spontaneously broken time reversal symmetry.

The new element that emerges as the width of the strip is reduced is that we need to include a small, but finite amplitude for quasiparticle
tunneling between the two boundaries of the strip. These quasiparticle tunneling terms will generically split the four-fold ground state
degeneracy discussed above, as we now demonstrate.

To see that quasiparticle tunneling will split the ground state degeneracy, it is sufficient to give one example. We consider the following
tunneling term:
\begin{equation}
 H_1 = \epsilon\int_0^L dx \cos(4\phi_1+4\phi_2+4\phi_3+4\phi_4+2\phi_5+2\phi_6). \label{finitewidth1}
\end{equation}
This term describes a process in which the quasiparticle $l^T = (4,4)$ tunnels from one boundary to another, while simultaneously two electron
spins are flipped at boundary $b$. Importantly, $H_1$ is charge conserving and time reversal invariant according to (\ref{trtrans}),
(\ref{trtrans1}), (\ref{trtrans2}), so there is no symmetry principle that prevents it from appearing in our Hamiltonian as the width of the
strip is reduced. Furthermore, from the expression (\ref{s2def}), we can see that $H_1 \sim (S_0^{\rm II})^2 + h.c.$. Therefore, by (\ref{mrep}),
the matrix elements of $H_1$ between the different ground states are of the form
\begin{equation}
 H_1 \sim \left( \begin{array} {cccc}
\epsilon  & 0 & 0 & 0 \\
0 & -\epsilon  & 0 & 0 \\
0 & 0 & \epsilon  & 0 \\
0 & 0 & 0 & -\epsilon
\end{array} \right).
\end{equation}
From these matrix elements, we see that in lowest order of perturbation theory, $H_1$ will split the four-fold ground state degeneracy into a two-fold degeneracy.

The discussion above establishes that as the width of the strip is reduced, the degeneracy will generically split from $4 \rightarrow 2$ as a
result of quasiparticle tunneling. The next question is whether it is possible for the degeneracy to split further. We now show that such
further splitting is not possible as long as time reversal symmetry is preserved. To establish this point, we will show that the matrix elements of local time reversal invariant perturbations $H_1$ between different ground states are always of the form
\begin{equation}
 H_1 =\left( \begin{array} {cccc}
\epsilon_1 & 0 & 0 & 0 \\
0 & \epsilon_2 & 0 & 0 \\
0 & 0 & \epsilon_1 & 0 \\
0 & 0 & 0 & \epsilon_2
\end{array} \right) \label{general-h1}
\end{equation}
with $\epsilon_1$ and $\epsilon_2$ being real numbers. It will then follow that the perturbation $H_1$ cannot split the two-fold degeneracy,
at least in lowest order of perturbation theory.

To see that $H_1$ has the form (\ref{general-h1}),  note $H_1$ must satisfy two requirements.
First, it must be time reversal invariant so that
\begin{equation}
 T H_1^* T =  H_1,
\label{trreq}
\end{equation}
where $T$ is given by (\ref{trep}). Second, $H_1$ must be local, that is $H_1 = \sum_a \mathcal H_{a}$ where each $\mathcal H_{a}$ acts on a finite region whose size is much smaller than
$L$. Using this fact, we can show that $\mathcal H_{a}$ commutes with $S_0^{\rm II}$ within the ground state subspace. Indeed, if $\gamma$ is the path along which $S_0^{\rm II}$ acts, we can always choose a new path $\gamma'$ that has no overlap with the region of support of $\mathcal H_{a}$.
Using path independence (\ref{pathind}) we can then deduce:
\begin{eqnarray}
\<m|\mathcal H_{a} S_{0\gamma}^{\rm II}|n\> &=&  \<m|\mathcal H_{a} S_{0\gamma'}^{\rm II}|n\> \nonumber \\
&=& \<m|S_{0\gamma'}^{\rm II}\mathcal  H_{a}|n\> \nonumber \\
&=& \<m|S_{0 \gamma}^{\rm II} \mathcal  H_{a}|n\>.
\label{commarg}
\end{eqnarray}
Hence
\begin{equation}
 \<m|S_{0\gamma}^{\rm II} H_1|n\> = \<m| H_1 S_{0\gamma}^{\rm II}|n\>.
\label{commreq}
\end{equation}
Equations (\ref{trreq}) and (\ref{commreq}) together imply that $ H_1$ has the form (\ref{general-h1}). Thus, we see a two-fold degeneracy is guaranteed by time reversal symmetry.

Note that if we include corrections due to finite length $L$ of the edges, $H_1$ does not have to satisfy the second requirement (\ref{commreq}). The two-fold degeneracy then will be lifted, leading to a unique ground state. This is easy to understand since spontaneous symmetry breaking is
impossible in finite-size systems.

\section{Local order parameter}
\label{localorderparam}

In Sec.~\ref{example} and Sec.~\ref{orderparameter}, we argued that the weak breaking of time reversal symmetry could be detected by
a nonlocal order parameter $W$ (\ref{string-example}). In this section, we study the properties of \emph{local} order parameters.
We show that local order parameters can also detect the symmetry breaking, but only very weakly. More precisely, we show that
the expectation values of local order parameters are exponentially small in the width of the strip.

The argument for the exponential suppression of local order parameters is as follows. Consider a local observable $O$ that is odd under time reversal, and that may live anywhere in the system, not restricted to the boundaries. Since $O$ is odd under time reversal, it follows from equation (\ref{trep}) that
\begin{equation}
\langle n|O|n\rangle =-\langle n+2|O|n+2\rangle.
\end{equation}
At the same time, since the ground state degeneracy is topologically protected, we know that the matrix elements of the local operator $O$ must take
the form
\begin{equation}
\<m|O| n\> = C\delta_{mn} + \epsilon_{mn},
\label{oeq}
\end{equation}
where $C$ is a constant and $\epsilon_{mn}$ is a finite-size correction that vanishes in the thermodynamic limit. The correction $\epsilon_{mn}$ is directly related to the finite-size splitting between topologically degenerate ground states and by analogy with this splitting, we expect $\epsilon_{mn}$ to depend exponentially on the width of the strip: $\epsilon_{mn} = \mathcal O(e^{-\alpha D})$. Combining these two equations, we conclude that
\begin{equation}
\<m|O|n\> =\mathcal O(e^{- \alpha D}).
\end{equation}
In other words, the expectation value of any local order parameter is exponentially small in the width of the strip.

The above argument shows that the expectation values of local order parameters are \emph{at most} exponentially small in the width of the strip.
In fact, we will now argue that local order parameters generically saturate this bound.
One way to see this is to consider the effects of quasiparticle tunneling. Consider, for example, the following term:
\begin{equation}
H_1=\epsilon \int dx \sigma(x) W(x).
\label{perturb}
\end{equation}
Here $\sigma(x) = \cos(\phi_5 + \phi_6)$ and $W(x)$ is the nonlocal order parameter given in (\ref{string-example}). The above term (\ref{perturb})
will generically appear in our edge Hamiltonian since it is time reversal invariant and charge conserving and therefore there is no symmetry prohibiting it. On the other hand, we expect the coefficient $\epsilon$ to be exponentially suppressed, $\epsilon\sim \exp(-\alpha D)$, since this term involves a tunneling process across the strip.

Including (\ref{perturb}) into our edge Hamiltonian has an important effect: it is not hard to see that this term induces a small expectation value for $\sigma$ since $W(x)$ obtains an expectation value in the unperturbed ground state. In fact, in lowest order of perturbation theory we have:
\begin{equation}
\<\sigma\> \sim \epsilon \cdot \<W\> \sim \exp(-\alpha D).
\label{sigeq}
\end{equation}
With equation (\ref{sigeq}), the argument is now complete: we have demonstrated that $\sigma$ -- a \emph{local} order parameter -- generically has an
exponentially small expectation value as a result of the finite width of the strip. Clearly, the same reasoning applies to other local order parameters as well.

In summary, we have shown that for local order parameters, the symmetry breaking is indeed weak -- exponentially suppressed by the width of the
annular strip. As we approach the narrow-strip limit, local order parameters obtain larger and larger expectation values,
and thereby the weak symmetry breaking becomes the usual ``strong'' symmetry breaking.

\section{Generalization}
\label{general}

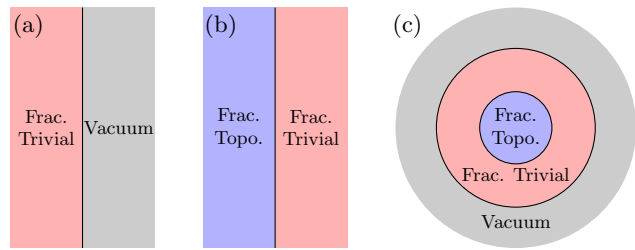
\begin{figure}[b]
\centering
\begin{tikzpicture}[scale=1.6]
\fill [red!30](0,0) rectangle +(-0.6,2);
\fill [black!20] (0,0) rectangle +(0.6,2);
\draw (0,0)--(0,2);
\node at (-0.3, 1)[align=center, scale=0.8]{Frac. \\Trivial};
\node at (0.3, 1)[align=center, scale=0.8]{Vacuum};
\node at (-0.45, 1.85){(a)};

\def\px {1.6}

\fill [blue!30](0+\px,0) rectangle +(-0.6,2);
\fill [red!30] (0+\px,0) rectangle +(0.6,2);
\draw (0+\px,0)--(0+\px,2);
\node at (0.3+\px, 1)[align=center, scale=0.8]{Frac. \\Trivial};
\node at (-0.3+\px, 1)[align=center, scale=0.8]{Frac. \\ Topo.};
\node at (-0.45+\px, 1.85){(b)};

\def\ppx {3.6}
\def\ppy {1}

\fill [black!20](0+\ppx,0+\ppy)  circle(1);
\draw [fill=red!30](0+\ppx,0+\ppy) circle(0.66);
\draw [fill=blue!30](0+\ppx,0+\ppy) circle(0.3);
\node at(0+\ppx,0+\ppy)[align=center, scale=0.8]{Frac. \\Topo.};
\node at(0+\ppx, -0.39+\ppy)[scale=0.8]{Frac. Trivial};
\node at(0+\ppx, -0.78+\ppy)[scale=0.8]{Vacuum};
\node at (-0.9+\ppx, .85+\ppy){(c)};
\end{tikzpicture}
\caption{The same geometries as in Fig.~\ref{fig1}, with the topological insulator replaced by a general fractional topological insulator  and the strong pairing insulator replaced by a general fractional trivial insulator.}
\label{fig6}
\end{figure}

In the above sections, we studied the paradox and the weak breaking of time reversal symmetry through the example in Fig.~\ref{fig1}. In this section, we construct many other similar examples, and show that these interesting phenomena are quite general. We consider the geometries in Fig.~\ref{fig6}, which is almost the same as Fig.~\ref{fig1} but the topological insulator is replaced by a general fractional topological insulator and the strong pairing insulator is replaced by a general fractional trivial insulator. We show that there exist many fractional topological insulators and fractional trivial insulators such that all boundaries in Fig.~\ref{fig6} are gapped without locally breaking any symmetry. Therefore, similarly to Fig.~\ref{fig1}c, a paradox occurs in Fig.~\ref{fig6}c: thinking of the annular strip as a wide boundary between the fractional topological insulator and the vacuum, we have gapped this boundary without apparently breaking any symmetry. At this point, we should not be surprised of the paradox, because it may be related to weak symmetry breaking. Indeed, we will show that in all the cases studied here, there is weak breaking of time reversal symmetry when the paradox occurs.

\subsection{The stability criterion}
\label{review}

To simplify our analysis of the stability of the boundaries in Fig.~\ref{fig6} and avoid similar technical analyses to those in Sec. \ref{example}, we will use the general edge stability criterion derived in Ref.~\onlinecite{levin12}. The criterion says that the boundary of an Abelian insulator, modelled by a $K$-matrix $\mathcal K$, carries a protected gapless edge mode if and only if the quantity $\frac{1}{e^*}\chi^T\mathcal{K}^{-1}\tau$ is odd. Here, $e^*$ is the smallest charge in the insulator,  $\chi$ is related to time reversal symmetry, and $\tau$ is related to charge conservation symmetry. Before using the criterion, we explain the meaning of the parameters $\tau$, $e^*$ and $\chi$.

As argued in Ref.~\onlinecite{levin12}, the boundary of a general time reversal invariant Abelian insulator is modeled by a $K$-matrix theory
\begin{align}
L = & \frac{1}{4\pi} (\mathcal K_{IJ}\partial_t\Phi_I \partial_x \Phi_J - \mathcal V_{IJ} \partial_x\Phi_I\partial_x\Phi_J),
\end{align}
where $\mathcal K$ and $\mathcal V$ are $2N\times 2N$ matrices. The parameters $\chi$ and $\tau$ tells us how the fields $\Phi_I$ transform under the basic symmetry transformations, i.e., the $U(1)$ charge symmetry and time reversal symmetry. Under a $U(1)$ charge transformation $\mathcal U(\theta)$ with $\theta$ the rotational angle, the fields transform as
\begin{equation}
\Phi_I \rightarrow \Phi_I + \theta \mathcal K^{-1}_{IJ} \tau_J,
\end{equation}
where $\tau$ is an $2N$-component integer vector. This expression defines $\tau$, which is called the ``charge vector''. According to this transformation, it is easy to check that the operator $e^{il^T\Phi}$ creates a quasiparticle with charge $l^T\mathcal K^{-1}\tau$. Then, the smallest charge $e^*$ can be defined as
\begin{equation}
e^{*}= {\rm min}_{l}(l^T \mathcal K^{-1}\tau).
\end{equation}

The time reversal symmetry $\mathcal T$  acting on the fields $\Phi_I$ leads to a transformation
\begin{equation}
\Phi_I \rightarrow T_{IJ}\Phi_J + \pi \mathcal K^{-1}_{IJ}\chi_J, \label{tr_transformation}
\end{equation}
where $T$ is a $2N\times 2N$ integer matrix, and $\chi$ is a $2N$-component integer vector. This expression defines $\chi$, which we can think  as a ``time reversal vector''. The parameters $\mathcal K$, $\tau$ cannot be chosen arbitrarily: As shown in Ref.~\onlinecite{levin12}, the requirement of time reversal invariance gives
\begin{align}
T^T\mathcal KT & =-\mathcal K, \nonumber\\
T\tau & = \tau. \label{tr-chargevector}
\end{align}
Also, the property of the time reversal symmetry, $\mathcal T^2=(-1)^{N_e}$, where $N_e$ is the total number of electrons in the system, gives the following constraints on $T$ and $\chi$:
\begin{align}
T^2 &=1,\nonumber\\
(1-T^T)\chi & \equiv \tau \,({\rm mod}\, 2). \label{requirement}
\end{align}
We see that the parameters $(\mathcal K, \tau, T, \chi)$ capture the essence of time reversal invariant Abelian insulators.

\subsection{Boundary stability in Fig.~\ref{fig6}a and Fig.~\ref{fig6}b}
\label{stability}

We now apply the stability criterion to the boundaries in Fig.~\ref{fig6}a and Fig.~\ref{fig6}b. Let us parameterize the fractional trivial insulator with ($\mathcal K_1, \tau_1, T_1, \chi_1$), and the fractional topological insulator with ($\mathcal K_2, \tau_2, T_2, \chi_2$). According to the very definitions of the fractional topological/trivial insulator, we must have $\frac{1}{e^*_1}\chi^T_1\mathcal K^{-1}_1\tau_1$ being even and $\frac{1}{e^*_2} \chi_2^T\mathcal K_2^{-1}\tau_2$ being odd, with $e^*_1$ and $e_2^*$ the smallest charges in the fractional trivial insulator and the fractional topological insulator respectively. Also, by construction, the boundary in Fig.~\ref{fig6}a can be gapped without breaking any symmetry.

The stability of the boundary in Fig.~\ref{fig6}b is not as straightforward. The boundary contains the edge modes from the central disk and the modes from the inner edge of the annular strip. Hence, the modes can be modeled by the $K$-matrix theory with parameters
\begin{align}
 \mathcal K_b=\left(
\begin{array}{cc}
-\mathcal K_1 & 0\\
0 &\mathcal K_2
\end{array}
\right)
,\quad \tau_b=\left(
\begin{array}{c}
\tau_1\\
\tau_2
\end{array}
\right)\nonumber\\
 T_b=\left(
\begin{array}{cc}
 T_1 & 0\\
0 & T_2
\end{array}
\right)
,\quad \chi_b=\left(
\begin{array}{c}
\chi_1\\
\chi_2
\end{array}
\right) \label{boundaryb-param}
\end{align}
The minus sign before $\mathcal K_1$ comes from the fact that the inner edge modes of the annular strip have opposite chirality to the outer edge modes.

Applying the stability criterion explained above, we see that the boundary in Fig.~\ref{fig6}b contains gapless edge modes, if and only if the quantity
\begin{equation}
\frac{1}{e_b^*}\chi_b^T{\mathcal K}_b^{-1}\tau_b=\frac{1}{ e_b^*}( \chi_2^T\mathcal K_2^{-1}\tau_2-\chi_1^T\mathcal K_1^{-1}\tau_1 ) \label{criteria_total}
\end{equation}
is odd. Here, $e_b^*$ is defined as
\begin{align}
e_b^* &={\rm min}_l( l^T{\mathcal K}_b^{-1}\tau_b)\nonumber \\\label{estar_total}
& ={\rm min}_{\{m,n\}}(m e_1^*+n e_2^*),
\end{align}
where $l$ is an integer vector, and $m$, $n$ are integer numbers. With this criterion, it will be easy for us to construct many examples where the boundary in Fig.~\ref{fig6}b is unstable. Stable boundaries also exist, however they are not interesting to us.

Before we construct other examples in which the boundary in Fig.~\ref{fig6}b is unstable, let us check that the criterion is consistent with the example from Fig.~\ref{fig1}. In that case, the strong pairing insulator has parameters
\begin{align}
 \mathcal K_1 & =\left(
\begin{array}{cc}
 8 & 0\\
0 & -8
\end{array}
\right)
,\quad \tau_1=\left(
\begin{array}{c}
2\\
2
\end{array}
\right)\nonumber \\
 T_1 & =\left(
\begin{array}{cc}
0 & 1\\
1 & 0
\end{array}
\right)
,\quad \chi_1 =\left(
\begin{array}{c}
0\\
0
\end{array}
\right).
\end{align}
and the topological insulator has parameters
\begin{align}
 \mathcal K_2 & =\left(
\begin{array}{cc}
 1 & 0\\
0 & -1
\end{array}
\right)
,\quad \tau_2=\left(
\begin{array}{c}
1\\
1
\end{array}
\right)\nonumber \\
 T_2 & =\left(
\begin{array}{cc}
0 & 1\\
1 & 0
\end{array}
\right)
,\quad \chi_2 =\left(
\begin{array}{c}
0\\
1
\end{array}
\right).
\end{align}
where $T_1$, $\chi_1$, $T_2$ and $\chi_2$ can be read off from the time reversal transformations (\ref{trtrans1}) and (\ref{trtrans2}). The parameters ($\mathcal K_b, \tau_b, T_b, \chi_b$) can be obtained according to (\ref{boundaryb-param}). Then, the smallest charges are $e_1^*=1/4$, $e_2^*=1$ and  $e^*_b=1/4$, and thereby $\frac{1}{e_1^*}\chi_1^T{\mathcal K}_1^{-1}\tau_1=0$ and $\frac{1}{e_b^*}\chi_b^T{\mathcal K}_b^{-1}\tau_b=-2$. Therefore, both boundaries in Fig.~\ref{fig1}a and Fig.~\ref{fig1}b are unstable, in agreement with the results  from Sec.~\ref{example}.

\subsection{Infinitely many examples of unstable boundary in Fig.~\ref{fig6}b}

Now we look for other examples in which the boundary in Fig.~\ref{fig6}b is unstable. In fact, we will show that there are infinitely many such examples.

We will restrict to a special class of Abelian insulators, which we call ``Abelian spin Hall insulators''. We will see below that this class is already enough for us to find infinitely many examples where the boundary in Fig.~\ref{fig6}b is unstable. Of course, there are also examples beyond Abelian spin Hall insulators, however we will not list them here.

The Abelian spin Hall insulators\cite{neupert11,levin12} has the parameters of the form
\begin{align}
\mathcal K &= \left(
\begin{array}{cc}
K & W \\
W^T & -K
\end{array}
\right),\quad
\tau= \left(
\begin{array}{c}
t \\
t
\end{array}
\right)\nonumber\\
T &= \left(
\begin{array}{cc}
0 & 1 \\
1 & 0
\end{array}
\right),\quad
\chi= \left(
\begin{array}{c}
0 \\
t
\end{array}
\right),
\end{align}
where $K$ is symmetric and $W$ is skew symmetric. One may check that the requirements in (\ref{tr-chargevector}) and (\ref{requirement}) are satisfied. We assume ${\rm gcd}(t)=1$ for simplicity. Let the fractional topological insulator and fractional trivial insulator in Fig.~\ref{fig6} both be of this type. In this case, the stability criterion (\ref{criteria_total}) simplifies further. As we explain below, the boundary stability of Fig.~\ref{fig6}b is  fully determined by the single number $1/e^*_1$: if it is even, the boundary is unstable; if it is odd, the boundary is protected. Using this result, it is then easy to construct  infinitely many examples in which the boundary in Fig.~\ref{fig6}b is unstable. For example, the following Abelian spin Hall insulators are fractional trivial insulators and all have their $1/e^*$ being even:
\begin{equation}
 K = \left(
\begin{array}{cc}
m & n\\
n & m
\end{array}
\right), \quad W=0,\quad
t= \left(
\begin{array}{c}
1 \\
1
\end{array}
\right).
\end{equation}
Here, $m$ and $n$ both are odd integers. The smallest charge is $e^*=1/(m+n)$. So, clearly $1/e^*$ is even.

The proof of the above result relies on a theorem of Abelian spin Hall insulators, proven in Appendix \ref{append_a}. The theorem states that the smallest charge in an Abelian spin Hall insulator with protected edge modes must have an odd denominator. On the other hand, the smallest charge of an Abelian spin Hall insulator without protected edge modes can have either odd or even denominator. Given this result, we see that  $1/e^*_2$ must be odd while $1/e^*_1$ can be either even or odd. Then, using the fact that $\frac{1}{e^*_1}\chi^T_1\mathcal K^{-1}_1\tau_1$ is even and $\frac{1}{e^*_2} \chi_2^T\mathcal K_2^{-1}\tau_2$ is odd, one finds that $\frac{1}{ e_b^*}\chi_b^T{\mathcal K}_b^{-1}\tau_b$, defined in (\ref{criteria_total}),  has the same parity as $1/e^*_1$. Hence, the boundary in Fig.~\ref{fig6}b can be gapped without breaking any symmetry if and only if $1/e_1^*$ is even.

\subsection{Weak symmetry breaking in Fig.~\ref{fig6}c}

\label{general3}

Let us now consider the geometry in Fig.~\ref{fig6}c. In the case that the boundaries $a$ and $b$ both are unstable, a similar paradox to that in Sec.~\ref{paradox} occurs: thinking of the annular strip as a wide edge, we can construct a gapped edge of a fractional topological insulator. Furthermore, according to the above subsection, there are infinitely many examples where this paradox arises.

The question then is: Do all these examples exhibit weak symmetry breaking, just like the topological insulator/strong pairing insulator example? Below we answer this question in the affirmative. We prove the answer by finding a nonlocal order parameter for time reversal symmetry breaking on a general ground. The existence of the nonlocal order parameter establishes the existence of weak breaking of time reversal symmetry. We will not discuss the ground state degeneracy here, though we expect it behaves similarly to the topological insulator/strong paring insulator example.

Let us set up some notations before searching for a nonlocal order parameter. We have already associated the fractional trivial insulator with parameters $(\mathcal K_1, \tau_1, T_1, \chi_1)$, and the fractional topological insulator with parameters $(\mathcal K_2, \tau_2, T_2, \chi_2)$. Let $\mathcal K_1$ be a $2N_1\times 2N_1$ matrix and $\mathcal K_2$ be a $2N_2 \times 2N_2$ matrix. Boundary $a$ is described by parameters $(\mathcal K_a, \tau_a, T_a, \chi_a)=(\mathcal K_1, \tau_1, T_1, \chi_1)$, and boundary $b$ is described by $(\mathcal K_b, \tau_b, T_b, \chi_b)$ defined in (\ref{boundaryb-param}). Let the edge modes of the annular strip at boundary $a$ be described by $\Phi_1$, the modes of the annular strip at boundary $b$ be described by $\Phi_1'$, and the modes of the disk be described by $\Phi_2$. In total, boundary $b$ is described by
\begin{equation}
\Phi_b=\left(
\begin{matrix}
\Phi_1' \\
\Phi_2
\end{matrix}
\right).
\end{equation}
For notational consistency, we also denote $\Phi_a=\Phi_1$.

According to Ref.~\onlinecite{levin12}, if a boundary is unstable, it is always possible to find simple scattering terms that preserve the symmetries to gap it out, just like what we have done in the example in Sec.~\ref{example}. Let us assume as in Ref.~\onlinecite{levin12} that the boundary $a$ is gapped by $N_1$ scattering terms
\begin{align}
U_{a,i}(x) \cos(\Lambda_{a,i}^T\mathcal K_a\Phi_a-\alpha_{a,i}(x)), \label{pert-a-general}
\end{align}
where $i=1, \dots, N_1$. The integer vector set $\{\Lambda_{a,i}\}$ is assumed to satisfy the following properties. First, they are linearly independent and neutral $\Lambda_{a,i}\tau_a=0$, and they satisfy the null vector criterion $\Lambda_{a, i}^T\mathcal K \Lambda_{a,j}=0$. Second, if a vector $\Lambda_{a,i}$ belongs to the set, its time reversal partner $-T\Lambda_{a,i}$ also belongs to the set. Finally, the vectors $\{\Lambda_{a,i}\}$ satisfy  the primitivity condition. These properties ensure that the scattering terms are charge conserving, time reversal invariant, and can gap the boundary $a$ without locally breaking any symmetry. Let the boundary $b$ be gapped by the scattering terms
\begin{align}
U_{b,i}(x) \cos(\Lambda_{b,i}^T\mathcal K_b\Phi_b-\alpha_{b,i}(x)), \label{pert-b-general}
\end{align}
with $i=1, \dots, (N_1+N_2)$. Again, $\{\Lambda_{b,i}\}$ should satisfy the symmetry requirements as well as the null vector criterion and the primitivity condition.

Once the boundaries are gapped,  we claim that there exists a nonlocal order parameter for the time reversal symmetry breaking. In the edge theory, it is expressed as
\begin{equation}
W(x) = \cos[\frac{1}{e^*_2}\tau_1^T(\Phi_{1} + \Phi_1')+\frac{1}{e_2^*}\tau_2^T\Phi_2].\label{string-general}
\end{equation}
It describes a physical process where a quasiparticle labeled by $l=\tau_1/e^*_2$ tunnels from boundary $a$ to boundary $b$, and at the same time a quasiparticle labeled by $l'=\tau_2/e_2^*$ is created at the edge of the disk. Note that the quasiparticle $l'$ is local because $\tau_2/e_2^*$ can be written as $\mathcal K_2\Lambda$ with an integer vector $\Lambda$, as is a consequence of the definition of $e_2^*$.

We prove the claim below. First, we show that $W$ is odd under time reversal transformation. According to (\ref{tr_transformation}), we have
\begin{align}
\Phi_1 &\rightarrow T_1\Phi_1 + \pi \mathcal K_1^{-1}\chi_1 \nonumber \\
\Phi_1' & \rightarrow T_1\Phi_1' -\pi \mathcal K_1^{-1}\chi_1\nonumber \\
\Phi_2 & \rightarrow T_2\Phi_2 +\pi \mathcal K_2^{-1}\chi_2.
\end{align}
Then, with the property (\ref{tr-chargevector}), we have
\begin{equation}
W(x)\rightarrow (-1)^{\frac{1}{e^*_2} \tau_2^T\mathcal K_2^{-1}\chi_2}W(x).
\end{equation}
Since the disk is assumed to be a fractional topological insulator, the quantity $\frac{1}{e^*_2} \tau_2^T\mathcal K_2^{-1}\chi_2$ is odd. Hence, $W\rightarrow -W$ under the time reversal transformation, as claimed.

Second, we show that $W(x)$ indeed acquires a nonvanishing ground-state expectation value. To see this, we first write $W$ in a slightly different form,
\begin{equation}
W=\cos[\frac{1}{e^*_2}\tau_a^T\Phi_{a} +\frac{1}{e_2^*}\tau_b^T\Phi_b].
\end{equation}
In the large $U$ limit, the cosine terms in (\ref{pert-a-general}) and (\ref{pert-b-general}) lock the fields $\Lambda_{a,i}^T\mathcal K_a\Phi_a$ and $\Lambda_{b,i}^T\mathcal K_b\Phi_b$ to classical values. We will now argue that the field $\tau_a^T\Phi_a$ and $\tau_b^T\Phi_b$ also acquire classical values. The integer vectors $\{\Lambda_{a,i}\}$ satisfy the neutrality condition $\Lambda_{a, i}^T\tau_a=0$, and the null vector criterion $\Lambda_{a,i}^T\mathcal K_a\Lambda_{a, j} =0$. Let $\mathbb X$ be the space spanned by the vectors satisfying the $N_1$ linear equations \{$\Lambda_{a,i}^T\mathcal K_a x=0$\}, where $x$ is a $2N_1$ dimensional vector. Then, $\mathbb X$ has dimension $N_1$. The null vector condition tells us that \{$\Lambda_{a, i}$\} is a basis of $\mathbb X$. From the neutrality condition, we see the vector $\mathcal K_a^{-1} \tau_a$ is also a solution of the linear equations \{$\Lambda_{a,i}^T\mathcal K_a x=0$\}. Hence, $\mathcal K_a^{-1} \tau_a \in \mathbb X$, implying that $\mathcal K_a^{-1} \tau_a $ can be expressed as a linear combination of \{$\Lambda_{a, i}$\}. Therefore, the field $\tau_a^T\Phi_1$ can be written as a linear combination of the fields $\Lambda_{a,i}^T\mathcal K_a\Phi_a$ and acquires a classical value. Similarly, we can show that $\tau_b^T\Phi_b$ acquires a classical value. Hence, we achieve our conclusion that the nonlocal operator $W(x)$ in (\ref{string-general}) indeed obtains an expectation value when both boundaries $a$ and $b$ are gapped. In general, the expectation value is nonvanishing, because the phases $\alpha_{a,i}$ in (\ref{pert-a-general}) and $\alpha_{b,i}$ in (\ref{pert-b-general}) are arbitrary. We conclude that our system exhibits weak time reversal symmetry breaking when boundaries $a$ and $b$ are gapped.

\section{Conclusion}
\label{conclusion}

To summarize, we have shown that the strong pairing insulator has the interesting property that \emph{both} its boundary with the vacuum, and its
boundary with a topological insulator can be fully gapped without breaking either time reversal or charge conservation symmetry. This
result is closely connected with an apparent paradox. The paradox occurs when we consider a geometry in which a disk-like region containing a topological insulator is surrounded by an annular strip made of a strong pairing insulator, which is in turn surrounded by the vacuum. Naively, it appears that if we gap both boundaries, we can construct a fully gapped interface between a topological insulator and the vacuum that does not break any symmetries -- a contradiction.

The resolution of this paradox is that the above system spontaneously breaks time reversal symmetry in an unusual way which we call ``weak symmetry breaking.'' This weak symmetry breaking cannot be detected by any local order parameter if the annular strip is much wider than the microscopic correlation length. The only order parameters that can see the symmetry breaking are \emph{nonlocal} string-like operators that describe quasiparticle tunneling across the strip. In addition, we have shown that this unusual symmetry breaking is associated with an unusual ground state degeneracy: the system has a four-fold ground state degeneracy which is topologically protected rather than the expected two-fold symmetry-protected degeneracy. We have also generalized these discussions to fractional topological insulator/fractional trivial insulator/vacuum sandwich structures, where we found a large class of other systems exhibiting weak symmetry breaking.

In the course of our analysis, we have developed several tools that may be useful more generally. Most notably, we derived general formulas for
the ground state degeneracy of gapped edges (appendix \ref{degAppend}) in different geometries. These formulas take account of \emph{both} topological degeneracy and symmetry-breaking degeneracy in a unified framework.

The present work can be loosely grouped with recent studies of exotic phenomena at gapped boundaries of topological insulators. A well known example of such phenomena is the discovery of Majorana zero modes at the edge of a topological insulator\cite{fu08}. Another interesting example is the recent observation that surface states of 3D symmetry protected phases can be gapped by forming 2D intrinsic topological orders. For instance, the surface of a 3D bosonic topological insulator can form a $\mathbb Z_2$ topological phase\cite{vishwanath13}. More recently, it was also shown the surface state of a 3D fermionic topological insulator can realize non-Abelian topological order\cite{metlitski13, chen13b, bonderson13, wang13}. We see that gapped edges/surfaces of topological phases have rather rich physics. The weak symmetry breaking phenomenon identified in this work provides another example of this richness.

We envision several directions for future research. One direction is to consider symmetry protected topological phases with symmetries beyond charge conservation and time reversal symmetry. It would be interesting to see if the paradox and the associated weak symmetry breaking generalizes to
these systems. It would also be interesting to see if analogous phenomena occur at the surfaces of three dimensional topological insulators.

{\it Note added}. After this work was submitted for publication, we noticed a related work\cite{lu-lee} which also studies unconventional ways of gapping out the edges of 2D symmetry protected topological phases.

\begin{acknowledgments}
C.W. thanks M. Cheng, C.-H. Lin and A. Lobos for helpful discussions. C.W. and M.L. acknowledge support from Microsoft Q, JQI-NSF-PFC and the Alfred P. Sloan foundation.
\end{acknowledgments}

\appendix

\section{Degeneracy of gapped edges}
\label{degAppend}

In this appendix, we calculate the ground state degeneracy of general gapped edges modeled in the $K$-matrix formalism. The edges are gapped by backscattering terms, i.e., cosine potentials. Various geometries are considered, including a disk geometry, a cylindrical geometry and the geometry in Fig.~\ref{fig6}c. Let us first list the results, then prove them in the following subsections.

{\it Disk geometry}---We consider an edge of a disk described by a Lagrangian
\begin{align}
L&(\Phi, \mathcal K, \{\Lambda_i\})= \frac{1}{4\pi} \left(\partial_t\Phi^T \mathcal K\partial_x\Phi - \partial_x \Phi^T \mathcal V\partial_x\Phi\right) \nonumber \\ & + U_1\cos(\Lambda_1^T\mathcal K\Phi) + \dots + U_N \cos(\Lambda_N^T\mathcal K \Phi) \label{disk-lagrangian}
\end{align}
where $\mathcal K$ is a $2N\times 2N$ matrix and $\Phi$ contains $2N$ components. The $2N$ dimensional integer vectors $\{ \Lambda_i\}$  are linearly independent and satisfy the null vector criterion $ \Lambda_i^T \mathcal K \Lambda_j = 0$, so that the edge is gapped at large $U$'s. We will show that the ground state degeneracy (GSD) in this model is given by
\begin{equation}
{\rm GSD}_{\rm disk} = {\rm gcd}(\text{$N\times N$ minors of } \mathcal M) \label{diskdeg}
\end{equation}
where $\mathcal M = (\Lambda_1, \dots, \Lambda_N)$, a $2N\times N$ matrix, and $\rm gcd$ stands for ``greatest common divisor''.

{\it Cylindrical geometry}---In this case, there are two boundaries: the right boundary $a$ and the left boundary $b$. We consider a model in which the two boundaries are described by the edge Lagrangian
\begin{equation}
 L(\Phi_a,\mathcal K, \{\Lambda_{a,i}\}) + L(\Phi_b, -\mathcal K, \{\Lambda_{b,i}\})
\end{equation}
where the form of $L$ follows (\ref{disk-lagrangian}). Both fields $\Phi_a$ and $\Phi_b$ contain $2N$ components. $\{\Lambda_{a,i}\}$ are linearly independent $2N$ dimensional integer vectors, as are $\{\Lambda_{b,i}\}$. In addition, $\{\Lambda_{a,i}\}$ and $\{\Lambda_{b,i}\}$ satisfy the null vector criterion, $\Lambda_{a,i}^T\mathcal K \Lambda_{a,j} =\Lambda_{b,i}^T\mathcal K \Lambda_{b,j} =0$.  We will show that when the boundaries are gapped at large $U$'s, the ground state degeneracy of the system is given by
\begin{equation}
{\rm GSD}_{\rm cylinder} = {\rm gcd}(\text{$2N\times 2N$ minors of } \mathcal M)
\end{equation}
and $\mathcal M$ is a $4N\times 2N$ matrix
\begin{equation}
\mathcal M = \left(
\begin{matrix}
\mathcal K \Lambda_{a1} & \cdots & \mathcal K \Lambda_{aN} & 0 & \cdots & 0 \\
    \Lambda_{a1} & \cdots &  \Lambda_{aN} & \Lambda_{b1} & \cdots & \Lambda_{bN}
\end{matrix}
\right).
\end{equation}

{\it The geometry in Fig.~\ref{fig6}c}---Here we assume the two boundaries, the outer boundary $a$ and the inner boundary $b$, are modeled by the Lagrangian
\begin{equation}
L(\Phi_a, K_s, \{\Lambda_{a,i}\}) + L(\Phi_b, \mathcal K_b, \{\Lambda_{b,i}\}),
\end{equation}
where $\Phi_a$ describes $2N$ modes on boundary $a$, and $\Phi_b$ describes $2N+2M$ edge modes on boundary $b$ with $2N$ modes from the annular strip and $2M$ modes from the disk. The disk and the annular strip are associated with $K$-matrices $\mathcal K_d$ and $\mathcal K_s$ respectively, so $\mathcal K_b = {\rm diag}(-\mathcal K_s, \mathcal K_d)$. The integer vectors $\{\Lambda_{a,i}\}$ are $2N$ dimensional  and $\{\Lambda_{b,i}\}$ are $2N+2M$ dimensional. Both are linearly independent and satisfy the null vector criterion.  We will show that when the boundaries are gapped, the degeneracy in this geometry is given by
\begin{equation}
{\rm GSD}_{\rm Fig.\ref{fig6}c} = {\rm gcd}\left(\text{$(2N+M)\times (2N+M)$ minors of } \mathcal M\right),
\end{equation}
where $\mathcal M$ is a $(4N+2M)\times(2N+M)$ matrix. The first $N$ columns of $\mathcal M$ are
\begin{equation}
\left(
\begin{matrix}
\mathcal K_s\Lambda_{a,i}\\ \Lambda_{a,i}\\ 0_{2M}
\end{matrix}
\right), \quad i=1,\dots, N
\end{equation}
with $0_{2M}$ the $2M$ dimensional zero vector, and the last $M+N$ columns of $\mathcal M$ are
\begin{equation}
\left(
\begin{matrix}
0_{2N}\\ \Lambda_{b,i}
\end{matrix}
\right), \quad i=1,\dots, N+M
\end{equation}
with $0_{2N}$ the $2N$ dimensional zero vector.

\subsection{Disk geometry}

We will prove the formula (\ref{diskdeg}) by mapping the edge theory to a collection of non-chiral Luttinger liquids, for which the degeneracy is easily seen. We illustrate our approach with the example of the strong pairing insulator. We then prove the formula (\ref{diskdeg}) for general cases.

\subsubsection{Strong pairing insulator}

Consider the following model for the edge of a strong pairing insulator,
\begin{align}
L_{sp} &=\frac{8}{4\pi} \partial_x \phi_1 (\partial_t \phi_1 - v \partial_x \phi_1)\nonumber\\
& - \frac{8}{4\pi} \partial_x \phi_2 (\partial_t \phi_2 + v \partial_x \phi_2)\nonumber \\
& + U \cos(8\phi_1 + 8 \phi_2).
\end{align}
In this case, the matrix
\begin{equation}
\mathcal M  =\left(\begin{matrix}
1\\
-1
\end{matrix}
\right).
\end{equation}
So, according to the formula (\ref{diskdeg}), the ground state is nondegenerate. Below we will establish this result with a systematic calculation.

We will use a Hamiltonian formulation of the edge theory. The Hamiltonian is given by
\begin{equation}
H = \frac{8v}{4\pi} \left[ (\partial_x\phi_1)^2 + (\partial_x\phi_2)^2\right] -U \cos(8\phi_1 + 8 \phi_2),
\end{equation}
where the basic commutation relations are
\begin{align}
[\phi_1(x), \partial_y\phi_1(y)] &= \frac{i}{8}2\pi \delta(x-y), \nonumber \\
[\phi_2(x), \partial_y\phi_2(y)] &= -\frac{i}{8}2\pi \delta(x-y).
\end{align}
The particle densities are $\partial_x\phi_1/2\pi$ and $\partial_x \phi_2/2\pi$. On a disk geometry, the total number of particles in each component must be integers. Therefore, $H$ should be diagonalized with the constraints
\begin{align}
\int_0^L dx \ \frac{1}{2\pi}\partial_x\phi_1(x) = p_1 \ , \ \int_0^L dx \ \frac{1}{2\pi}\partial_x\phi_2(x) = p_2
\end{align}
where $p_1, p_2$ are integers, and $L$ is the length of the edge.

We wish to understand the ground state degeneracy of the Hamiltonian $H$ through a mapping to the standard non-chiral Luttinger liquid. To this end, we consider the change of variables
\begin{eqnarray}
\theta &=& 8 (\phi_1 + \phi_2), \nonumber \\
\varphi &=& \frac{1}{2} (\phi_1 - \phi_2).
\end{eqnarray}
In terms of these variables, we have
\begin{equation}
H = \frac{v}{4\pi} \left[\frac{(\partial_x\theta)^2}{16} + 16(\partial_x\varphi)^2\right]-U\cos(\theta)
\label{H0trans}
\end{equation}
with the commutation relations
\begin{align}
[\theta(x), \partial_y\varphi(y)] &= i2\pi\delta(x-y). \nonumber \\
[\theta(x), \partial_y\theta(y)] & =[\varphi(x), \partial_y\varphi(y)]=0.
\end{align}
The constraints become
\begin{align}
\int_0^L dx \ \frac{1}{2\pi} \partial_x\theta(x) & = 8(p_1+p_2) \ , \nonumber\\
 \ \int_0^L dx \ \frac{1}{2\pi}\partial_x\varphi(x) & = \frac{1}{2}(p_1 - p_2).
\label{constraintsp}
\end{align}

We observe that the above Hamiltonian (\ref{H0trans}) and commutation relations are identical to those of a non-chiral Luttinger liquid with a backscattering term
\begin{align}
H_{LL} = \pi( v_\theta \rho_\theta^2 + v_\varphi \rho_\varphi^2)-U\cos(\theta),\nonumber\\
[\rho_\theta(x), \rho_\varphi(y)] = \frac{i}{2\pi}\partial_x\delta(x-y),
\end{align}
where $v_\theta = v/16$ and $v_\varphi = 16 v$ and
\begin{align}
\rho_\theta = \frac{1}{2\pi} \partial_x \theta \ , \ \rho_\varphi = \frac{1}{2\pi} \partial_x \varphi.
\end{align}
Here, $\rho_\theta$ and $\rho_\varphi$ are the density of vortices and particles in the non-chiral Luttinger liquid.

On the other hand, the constraints (\ref{constraintsp}) are \emph{not} the same as a standard Luttinger liquid. Indeed, assuming the Luttinger liquid is defined with periodic boundary conditions, it will have the constraint that the number of vortices and the number of particles must be integer. That is,
\begin{align}
\int_0^L dx \rho_\theta(x) = p_\theta \ , \ \int_0^L dx \rho_\varphi(x) = p_\varphi
\label{constraintll}
\end{align}
where $p_\theta, p_\varphi$ are integers. These constraints are clearly different from (\ref{constraintsp}).

For this reason, $H$ is not {\it exactly} equivalent to a non-chiral Luttinger liquid, $H_{LL}$. However, we will now argue that they do have identical low energy spectra for large $U$. In particular, they have the same ground state degeneracy.

The first step is to note that the low energy eigenstates of $H$ all have $p_1 + p_2 = 0$.
Indeed, this follows immediately from the fact that
\begin{equation}
p_1 + p_2 = \frac{1}{2\pi} \int_0^L (\partial_x \phi_1 + \partial_x \phi_2)
\end{equation}
together with the fact that $\phi_1 + \phi_2$ is locked to the minimum of the cosine potential for all low energy states. By the same reasoning, we can see that the low energy eigenstates of $H_{LL}$ all have $p_\theta = 0$.

The next step is to note that there is a unitary equivalence between the Hamiltonian $H$
\emph{defined within the subspace $p_1 + p_2 = 0$}, and the Hamiltonian $H_{LL}$ \emph{defined with the subspace $p_\theta = 0$}. Indeed, we have already seen that the two Hamiltonians and commutation relations match up; all we have to check is that the constraints match as well. To this end, we note that within the $p_1 + p_2 = 0$ subspace, the constraints (\ref{constraintsp}) reduce to
\begin{align}
\int_0^L dx \rho_\theta(x) = 0 \ , \ \int_0^L dx \rho_\varphi(x) = \frac{1}{2}(p_1 - p_2) = \text{integer},
\end{align}
while within the $p_\theta = 0$ subspace, the constraints (\ref{constraintll}) reduce to
\begin{align}
\int_0^L dx \rho_\theta(x) = 0 \ , \ \int_0^L dx \rho_\varphi(x) = p_\varphi = \text{integer},
\end{align}
We can see that the constraints do in fact match.

Now that we have shown that the two Hamiltonians $H$ and $H_{LL}$ have identical low energy spectra. Since it is clear the $H_{LL}$ has a non-degenerate ground state, the Hamiltonian $H$ also has a non-degenerate ground state.

\subsubsection{General cases}

Now we consider a general edge on a disk, modeled by the Lagrangian
\begin{align}
L&(\Phi, \mathcal K, \{\Lambda_i\}) = \frac{1}{4\pi} \left(\partial_t\Phi^T \mathcal K\partial_x\Phi - \partial_x \Phi^T \mathcal V\partial_x\Phi\right) \nonumber \\ & + U_1\cos(\Lambda_1^T\mathcal K\Phi) + \dots + U_N \cos(\Lambda_N^T\mathcal K \Phi) \label{lagrangian}
\end{align}
where $\Phi$ is a $2N$-component field, $\mathcal K$ is a $2N\times 2N$ symmetric nonsingular integer matrix, and $\{\Lambda_i\}$ are $2N$ dimensional linearly independent integer vectors satisfying the null vector criterion, $\Lambda_i^T\mathcal K\Lambda_j=0$. The basic commutation relations are
\begin{equation}
[\Phi_I(x), \partial_y\Phi_J(y)] = i2\pi \mathcal K^{-1}_{IJ} \delta(x-y).
\end{equation}
The particle densities are $\partial_x\Phi_I(x)/2\pi$. On the disk geometry, the total number of particles of each component should be integers; thus we have the constraints
\begin{equation}
\int_0^L dx \ \frac{1}{2\pi} \partial_x\Phi_I = p_I \label{constraint-general}
\end{equation}
where $p_I$ are integers.

We would like to understand the ground state degeneracy of the Lagrangian (\ref{lagrangian}) with constraints (\ref{constraint-general}), by generalizing the discussion of the strong pairing insulator. We will map the model to a system of $N$ standard non-chiral Luttinger liquids. The three important things to keep track of during the mapping are: (1) the commutation relations, or equivalently, the $K$-matrix $\mathcal K$; (2) the form of cosine potentials, or equivalently, the corresponding integer vectors $\{\Lambda_i\}$; and (3) the constraints from total particle numbers. We will not keep track of the kinetic part of the Hamiltonian, since it is not important at the end when we take the large $U$ limit.

The mapping involves three steps. The first step is to simplify the problem by making use of the Smith normal form for integer matrices. Let $\mathcal M$ be the matrix with columns $\Lambda_1, \dots, \Lambda_N$. According to the Smith normal form, the matrix $\mathcal M$ can be written as $\mathcal M= S D R$. Here, $S$ is a $2N\times 2N$ integer matrix and $R$ is an  $N\times N$ integer matrix, both with determinant 1. $D$ is $2N\times N$ integer matrix of the form
\begin{equation}
D= \left(\begin{array}{cccc}
d_1 & 0 & \cdots & 0 \\
0 & d_2 & \cdots & 0 \\
\vdots & \vdots & \vdots & \vdots \\
0 & 0 & \cdots & d_N\\
\vdots & \vdots & \vdots & \vdots \\
0 & 0 & \cdots & 0
\end{array}
\right),
\end{equation}
where $d_I$ are integers. Since $\{\Lambda_i\}$ are linearly independent, the rank of the matrix $D$ is $N$, which means no $d_I$ is zero. Alternatively, we write $D$ as
\begin{equation}
D = \left(\begin{array}{c}
\bar D \\ {  0}
\end{array}\right),
\end{equation}
where $\bar D$ is a diagonal $N\times N$ integer matrix.

With the above Smith normal form of the matrix $\mathcal M$, we make the change of variables
\begin{equation}
\Phi' = S^{-1} \Phi. \label{transformation}
\end{equation}
Accordingly, the Lagrangian transforms as
\begin{align}
L(\Phi, \mathcal K, \{\Lambda_i\})\rightarrow L(\Phi', \mathcal K', \{\Lambda_i'\}) \nonumber
\end{align}
with
\begin{align}
\mathcal K' = S^T \mathcal K S, \nonumber \\
\Lambda'_i = S^{-1} \Lambda_i. \nonumber
\end{align}
The new parameters $\{\Lambda_i'\}$ and $\mathcal K'$ acquire simpler forms. The matrix
\begin{equation}
(\Lambda_1', \dots, \Lambda_N') = S^{-1}\mathcal M = \left(\begin{array}{c}
\bar D \\ {  0}
\end{array}\right)R= \left(\begin{array}{c}
\bar D R \\ {  0}
\end{array}\right),
\end{equation}
which means all $\Lambda'_i$ are vectors with the last $N$ components vanishing. According to the null vector criterion, $\mathcal M^T \mathcal K \mathcal M = 0$. Then, $R^T D^T \mathcal K' D R =0$, that is
\begin{equation}
(R^T \bar D, 0) \mathcal K' \left(\begin{array}{c}
\bar D R \\ {  0}
\end{array}\right)=0. \nonumber
\end{equation}
Therefore, $\mathcal K'$ must have the following form
\begin{equation}
\mathcal K' = \left(\begin{array}{cc}
0 & A\\
A^T & B
\end{array}\right),
\end{equation}
where $A$ is nonsingular and $B$ is symmetric.

In terms of the new variables $\Phi_I'$, the constraints (\ref{constraint-general}) become
\begin{equation}
\int_0^L dx \ \frac{1}{2\pi} \partial_x\Phi_I' = \sum_JS_{IJ}^{-1}p_J=p_I' .
\end{equation}
Since $S$ is an integer matrix with determinant 1, its inverse $S^{-1}$ is also an integer matrix with determinant $1$.  Moreover, all rows of $S^{-1}$ are primitive vectors. Therefore, the sum $\sum_JS_{IJ}^{-1}p_J$ generates all integers while varying the integers $p_J$. That is, $p_I'$ is an arbitrary integer.

The next step is to make another change of variables and map the problem to a problem of $N$ non-chiral Luttinger liquids. Let us make the change of variables
\begin{align}
\Phi''& = T \Phi'
\end{align}
where
\begin{equation}
\Phi''=\left(
\begin{matrix}
\theta_1\\
\vdots\\
\theta_N\\
\varphi_1\\
\vdots\\
\varphi_N
\end{matrix}
\right),\
T=\left(
\begin{matrix}
0 & A\\
1 & \frac{1}{2}(A^{-1})^TB
\end{matrix}
\right).
\end{equation}
It is easy to check that the new variables satisfy the commutation relations
\begin{align}
[\theta_I(x), \partial_y\theta_J(y)] &= [\varphi_I(x), \partial_y\varphi_J(y)] =0, \nonumber \\
[\theta_I(x), \partial_y\varphi_J(y)] &= i 2\pi \delta_{IJ}\delta(x-y).
\end{align}
In terms of the new variables, the cosine potentials are
\begin{equation}
-U_I\cos(\sum_J R_{JI}d_J\theta_J), \ I=1, \dots, N. \label{cos1}
\end{equation}
and the constraints are
\begin{align}
\int_0^L dx \frac{1}{2\pi} \partial_x\theta_I & = \sum_{J=1}^NA_{IJ}p'_{N+J},\nonumber\\
\int_0^L dx \frac{1}{2\pi} \partial_x\varphi_I & = p'_I + \frac{1}{2}\sum_{J, K =1}^N A^{-1}_{JI}B_{JK}p'_{N+K}. \label{constraint-g2}
\end{align}

We observe that the commutation relations match those of non-chiral Luttinger liquids. However, the constraints and cosine potentials are not ``standard''. In the standard Luttinger liquids, we expect cosine potentials to be
\begin{equation}
-U_I\cos( d_I \theta_I), \ I=1, \dots, N,  \label{cos2}
\end{equation}
and the constraints on total charges and vortices are
\begin{align}
\int_0^L dx \frac{1}{2\pi} \partial_x\theta_I & =p_{\theta I} \nonumber \\
\int_0^L dx \frac{1}{2\pi} \partial_x\varphi_I & = p_{\varphi I}. \label{constraint-g2-stand}
\end{align}
with $p_{\theta I}, p_{\varphi I}$ integers.

The final step is to take the large $U$ limit, and focus on the low energy spectra, as we did in the example of a strong paring insulator. In this limit, all the fields $\sum_JR_{JI}d_J\theta_J$ are locked at the minima of the cosine potentials. This means the low energy Hilbert space is constrained by
\begin{equation}
\sum_JR_{JI}d_J\theta_J = 2\pi s_I, \ I=1,\dots, N
\end{equation}
where $s_I$ are integers. Because $R$ is an integer matrix with determinant 1, we have
\begin{equation}
d_I\theta_I = 2\pi s_I', \ I=1,\dots, N
\end{equation}
where $s_I'$ are also arbitrary integers. One immediately realizes that the ``standard'' cosine potentials (\ref{cos2}) produce the same low-energy constraints.

The final thing to check is the constraints of total particle numbers. In the low energy subspace, $\theta_I$ are constants, so the constraints (\ref{constraint-g2}) become
\begin{equation}
\int_0^L dx \frac{1}{2\pi} \partial_x\theta_I = 0,\ \int_0^L dx \frac{1}{2\pi} \partial_x\varphi_I = p_I'.
\end{equation}
This matches the constraints (\ref{constraint-g2-stand}) of the non-chiral Luttinger liquids in the low energy subspace in which $p_{\theta I}=0$.

Now the model has been mapped to a standard Luttinger liquid problem, with the same set of constraints in the low-energy Hilbert space.  The ground state degeneracy of the system with $N$ standard non-chiral Luttinger liquids is easily seen. The $I$-th non-chiral Luttinger liquid has a degeneracy of $|d_I|$, so the overall degeneracy is $|d_1\cdots d_N|$. So the degeneracy of the original model (\ref{lagrangian}) is also $|d_1\cdots d_N|$.  Since $\mathcal M =S D R$ with $S$ and $R$ having determinant $1$,
\begin{align}
{\rm GSD_{disk}}&=|d_1\cdots d_N| \nonumber \\
 & = {\rm gcd}(\text{$N\times N$ minors of $\mathcal M$})
\end{align}
where $\mathcal M =(\Lambda_1, \dots, \Lambda_N)$.

\subsection{Cylindrical geometry}

In the cylindrical geometry, there are two edges: the right edge $a$ and the left edge $b$. We consider the case in which the two edges are well separated and individually gapped. The two edges are modeled by the Lagrangian
\begin{align}
 L_{\rm cylinder} =  L(\Phi_a, \mathcal K, \{\Lambda_{a,i}\}) +  L(\Phi_b, -\mathcal K, \{\Lambda_{b,i}\}) \label{cylinder-theory}
\end{align}
where the expression of the Lagrangian $L$ is given in (\ref{lagrangian}). (The velocity matrix $\mathcal V$ and scattering amplitudes $U_I$ are not important, so we do not keep track of them; all $U$'s are eventually taken to infinity.)  $\Phi_a$ and $\Phi_b$ are $2N$-component fields, describing edges $a$ and $b$ respectively. The integer vectors $\{\Lambda_{a,i}\}$ are linearly independent $2N$ dimensional vectors that satisfy the null vector criterion $\Lambda_{a,i}^T\mathcal K \Lambda_{a, j}=0$. The integer vectors $\{\Lambda_{b,i}\}$ are also linearly independent and satisfy the null vector criterion $\Lambda_{b,i}^T \mathcal K \Lambda_{b,j} = 0$. Note that $\{\Lambda_{a,i}\}$ and $\{\Lambda_{b,i}\}$ need not be equal, i.e., the two edges may be gapped in different ways.

The constraints from particle numbers are different from those for the disk geometry. In the cylindrical geometry, a single edge may contain some number of fractional particles, leading to the constraints
\begin{align}
\int_0^L dx \frac{1}{2\pi}\partial_x\Phi_{a,I} &  = \mathcal K_{IJ}^{-1}p_{aJ}, \nonumber \\
\int_0^L dx \frac{1}{2\pi}\partial_x\Phi_{b,I}  & = - \mathcal K_{IJ}^{-1}p_{bJ}, \label{cylinder-constraint1}
\end{align}
where $p_{aJ}, p_{bJ}$ are integers. However,  the total number of particles on the two edges must be integer, giving another constraint
\begin{equation}
\mathcal K_{IJ}^{-1}(p_{aJ}-p_{bJ})=q_I \label{cylinder-constraint2}
\end{equation}
where $q_I$ is an integer.

We would like to find the ground state degeneracy of the model (\ref{cylinder-theory}) with constraints (\ref{cylinder-constraint1}) and (\ref{cylinder-constraint2}), when the edges are gapped. Our strategy will be to map the model (\ref{cylinder-theory}) onto the model for the disk geometry, and then make use of the results obtained above.

First, let us write the theory (\ref{cylinder-theory}) in a compact form
\begin{equation}
L_{\rm cylinder}= L(\Phi, \mathcal K_t, \{\Lambda_{t,i}\})
\end{equation}
where
\begin{align}
\Phi & = \left(
\begin{matrix}
\Phi_a \\
\Phi_b
\end{matrix}
\right),\
\mathcal K_t=\left(
\begin{matrix}
\mathcal K & 0 \\
0 & -\mathcal K
\end{matrix}
\right), \nonumber \\
\Lambda_{t,i} & =\left(
\begin{matrix}
\Lambda_{a,i}\\
0
\end{matrix}
\right),\ \Lambda_{t,i+N}  =\left(
\begin{matrix}
0\\
\Lambda_{b,i}
\end{matrix}
\right),\nonumber
\end{align}
with $i=1,\dots, N$.

Next, we make the change of variables
\begin{equation}
\Phi'=\left(
\begin{matrix}
\Phi_a'\\
\Phi_b'
\end{matrix}
\right)
= S
\left(
\begin{matrix}
\Phi_a\\
\Phi_b
\end{matrix}
\right), \ S=\left(
\begin{matrix}
\mathcal K & 0\\
1 & 1
\end{matrix}
\right).\label{transformation2}
\end{equation}
Then, the Lagrangian transforms as $L(\Phi, \mathcal K_t, \{\Lambda_{t,i}\}) \rightarrow L(\Phi', \mathcal K_t', \{\Lambda_{t,i}'\})$, with the new parameters defined as $\mathcal K_t'  = (S^T)^{-1}\mathcal K_t S^{-1}$ and $\Lambda_{t,i}'  = S\Lambda_{t,i}$. The explicit expressions of the new parameters are
\begin{align}
\mathcal K_t' &  = \left(
\begin{matrix}
0 & 1\\
1 & -\mathcal K
\end{matrix}
\right), \nonumber \\
\Lambda_{t,i}'  & =\left(
\begin{matrix}
\mathcal K \Lambda_{a,i}\\
\Lambda_{a,i}
\end{matrix}
\right),\
\Lambda_{t,i+N}'  =\left(
\begin{matrix}
0\\
\Lambda_{b,i}
\end{matrix}
\right),
\end{align}
where $i=1, \dots, N$. Note that the null vector criterion is still satisfied by $\{\Lambda_{t,i}'\}$: $\Lambda_{t,i}'^T\mathcal K_t' \Lambda_{t, j}'=0.$

Finally, we check the constraints after the change of variables. In terms of the new variables, the constraints are
\begin{equation}
\int_0^L dx \frac{1}{2\pi}\partial_x\Phi_{aI}' =p_{aI}, \ \int_0^L dx \frac{1}{2\pi}\partial_x\Phi_{bI}' = q_I. \label{cylinder-constraint3}
\end{equation}
where $p_{aI}$ and $q_I$ are arbitrary integers.

We see that the new model for the edges on the cylindrical geometry, with the Lagrangian $L(\Phi', \mathcal K_t', \{\Lambda_{t,i}'\})$ and constraints (\ref{cylinder-constraint3}), is in the same form as the edge model for the disk geometry. So, we can make use of the results for the disk geometry to obtain the ground state degeneracy when the edges are gapped. According to the results in the above subsection, the ground state degeneracy is given by
\begin{equation}
{\rm GSD}_{\rm cylinder} = {\rm gcd}(\text{$2N\times 2N$ minors of } \mathcal M),
\end{equation}
where the matrix $\mathcal M = (\Lambda_{t1}', \dots, \Lambda_{t, 2N}')$, i.e.,
\begin{equation}
\mathcal M = \left(
\begin{matrix}
\mathcal K\Lambda_{a1} & \dots & \mathcal K\Lambda_{aN} & 0 & \dots &0\\
\Lambda_{a1} & \dots & \Lambda_{aN} & \Lambda_{b1} & \dots &\Lambda_{bN}
\end{matrix}
\right) .
\end{equation}

\subsection{The geometry Fig.~\ref{fig6}c in the main text}

The ground state degeneracy for the geometry in Fig.~\ref{fig6}c can be obtained in a way similar to that for the cylindrical geometry: again, we map the problem to the one in the disk geometry.

Let the annular strip be associated with a $2N\times 2N$ $K$-matrix $\mathcal K_s$, and the disk be associated with a $2M\times 2M$ $K$-matrix $\mathcal K_d$. The two boundaries, the outer boundary $a$ and inner boundary $b$,  are modeled by the Lagrangian
\begin{equation}
L_{\rm Fig.~\ref{fig6}c} = L(\Phi_a, \mathcal K_{s}, \{\Lambda_{a,i}\}) + \ L(\Phi_b, \mathcal K_b, \{\Lambda_{b,i}\}) \label{c-theory}
\end{equation}
where the expression for $L$ is given by (\ref{lagrangian}). $\Phi_a$ is a $2N$-component field and $\Phi_b$ is a $(2N+2M)$-component field. $\Phi_b$ is decomposed into two parts, a $2N$-component field $\Phi_{b1}$ describing the modes from the inner edge of the annular strip and a $2M$-component field $\Phi_{b2}$ describing the modes from the edge of the disk. At the same time, $\mathcal K_b$ can be written as
\begin{equation}
\mathcal K_b = \left(
\begin{matrix}
-\mathcal K_s & 0\\
0 & \mathcal K_d
\end{matrix}
\right).
\end{equation}
The integer vector sets $\{\Lambda_{a,i}\}$ and $\{\Lambda_{b,i}\}$ satisfy the null criterion individually, and vectors in each set are linearly independent. The constraints of particle numbers on each edge are
\begin{align}
\int_0^L dx \frac{1}{2\pi}\partial_x\Phi_{aI}& = \mathcal K_{s,IJ}^{-1}p_{aJ}, \nonumber \\
\int_0^L dx \frac{1}{2\pi}\partial_x\Phi_{b1,I}& = - \mathcal K_{s,IJ}^{-1}p_{b1,J},\nonumber \\
\int_0^L dx \frac{1}{2\pi}\partial_x\Phi_{b2,I} & = p_{b2,I}, \label{c-constraint1}
\end{align}
where $p_{aI}, p_{b1,I}, p_{b2,I}$ are integers. Like the cylindrical geometry, an additional constraint comes from the requirement that the total number of particles on the annular strip should be an integer:
\begin{equation}
\mathcal K_{s,IJ}^{-1}(p_{aJ}-p_{b1,J})=q_I \label{c-constraint2}
\end{equation}
where $q_I$ is an integer.

So, the boundaries in Fig.~\ref{fig6}c are modeled by the Lagrangian (\ref{c-theory}) with constraints (\ref{c-constraint1}) and (\ref{c-constraint2}). We now map this model to the one for the disk geometry, following the same steps for the cylindrical geometry. First, we write the Lagrangian in a compact form
\begin{equation}
L_{\rm Fig.~\ref{fig6}c}=L(\Phi, \mathcal K, \{\Lambda_{i}\})
\end{equation}
where
\begin{align}
\Phi & = \left(
\begin{matrix}
\Phi_a \\
\Phi_b
\end{matrix}
\right),\
\mathcal K=\left(
\begin{matrix}
\mathcal K_s & 0 \\
0 & \mathcal K_b
\end{matrix}
\right), \nonumber \\
\Lambda_{i} & =\left(
\begin{matrix}
\Lambda_{a,i}\\
0
\end{matrix}
\right),\ \Lambda_{j+N}  =\left(
\begin{matrix}
0\\
\Lambda_{b,j}
\end{matrix}
\right),\nonumber
\end{align}
with $i=1,\dots, N$ and $j=1, \dots, N+M$.

Then, we make the change of variables
\begin{equation}
\Phi'=\left(
\begin{matrix}
\Phi_a'  \\
\Phi_{b1}' \\
\Phi_{b2}'
\end{matrix}
\right) =
S \left(
\begin{matrix}
\Phi_a  \\
\Phi_{b1}\\
\Phi_{b2}
\end{matrix}
\right),\ S=
\left(
\begin{matrix}
\mathcal K_s & 0 & 0 \\
1 & 1 & 0 \\
0 & 0 & 1
\end{matrix}
\right).
\end{equation}
The Lagrangian changes accordingly, $L(\Phi, \mathcal K, \{\Lambda_{i}\})\rightarrow L(\Phi', \mathcal K', \{\Lambda_{i}'\})$ with the new parameters
\begin{align}
\mathcal K' &= \left(\begin{matrix}
0 & 1 & 0\\
1 & -\mathcal K_s & 0\\
0 & 0 & \mathcal K_d
\end{matrix}
\right), \nonumber \\
\Lambda_i' & = \left(\begin{matrix}
\mathcal K_s\Lambda_{a,i} \\
\Lambda_{a,i} \\
0_{2M}
\end{matrix}
\right), \ \Lambda_{j+N}' = \left(\begin{matrix}
0_{2N} \\
\Lambda_{b,j}
\end{matrix}
\right),  \label{c-newpara}
\end{align}
where $i=1,\dots, N$, $j=1, \dots, N+M$, and $0_{2N}$ and $0_{2M}$ are the $2N$- and $2M$-dimensional zero vectors respectively. Note that $\Lambda_{a,i}$ are $2N$-dimensional, $\Lambda_{b,i}$ are $(2M+2N)$-dimensional.

In terms of the new variables, the constraints become
\begin{align}
\int_0^L dx \frac{1}{2\pi}\partial_x\Phi_{aI}' &= p_{aI}, \nonumber\\
\int_0^L dx \frac{1}{2\pi}\partial_x\Phi_{b1,I}' &= q_I,\nonumber \\
\int_0^L dx \frac{1}{2\pi}\partial_x\Phi_{b2,I}' &= p_{b2,I} \label{c-constraint3}
\end{align}
where $p_{aI}, q_I, p_{b2,I}$ are arbitrary integers.

Therefore, in terms of the new variables, the boundaries in Fig.~\ref{fig6}c are modeled by the Lagrangian $L(\Phi', \mathcal K', \{\Lambda_{i}'\})$ with constraints (\ref{c-constraint3}), which is in the same form as the model for the disk geometry. Thus, according to the results for the disk geometry, the degeneracy is given by
\begin{equation}
{\rm GSD}_{\rm Fig.\ref{fig6}c} = {\rm gcd}[\text{$(2N+M)\times (2N+M)$ minors of } \mathcal M],
\end{equation}
with the matrix
\begin{equation}
\mathcal M = \left( \Lambda_1', \dots, \Lambda_{2N+M}'
\right),
\end{equation}
where $\{\Lambda_{i}'\}$ are given in (\ref{c-newpara}).

\section{A theorem about Abelian spin Hall insulators}
\label{append_a}
In this appendix, we prove a theorem about Abelian spin Hall insulators, i.e., those with parameters
\begin{align}
\mathcal K &= \left(
\begin{array}{cc}
K & W \\
W^T & -K
\end{array}
\right),\quad
\tau= \left(
\begin{array}{c}
t \\
t
\end{array}
\right),\nonumber\\
T &= \left(
\begin{array}{cc}
0 & 1 \\
1 & 0
\end{array}
\right),\quad
\chi= \left(
\begin{array}{c}
0 \\
t
\end{array}
\right),
\label{a1}
\end{align}
where $K$ is symmetric and $W$ is skew symmetric. Before we state the theorem, we make two comments. First, we assume that ${\rm gcd}(\tau)=1$ for simplicity. With this assumption, the inverse of the smallest charge $e^*$, defined as $e^*={\rm min}_l(l^T\mathcal K^{-1}\tau)$, is always an integer. (In fact, even if we start with ${\rm gcd}(\tau)\neq 1 $, it is possible to find an equivalent description with ${\rm gcd}(\tau)=1$. For example, a description with $\mathcal K' ={\rm diag }(\mathcal K, 1, -1, -1, 1)$ and $\tau' = (\tau, 1, 1,1 ,1)$ is equivalent to the description with $\mathcal K$ and $\tau$, because the two descriptions result in the same fractional statistics between quasiparticles. Clearly, ${\rm gcd}(\tau')=1$.) Second, we consider only fermionic insulators in which the constituent particles are electrons; in these systems, fermionic excitations must have odd charge while bosonic excitations must have even charge. This requirement leads to a constraint on $K$ and $t$, that is $ K_{II} \equiv t_I \, ({\rm mod}\,2)$.

The theorem is:
\begin{thm}
For Abelian spin Hall insulators, if $1/e^*$ is even, the quantity $\frac{1}{e^*} \chi^T\mathcal K^{-1} \tau$ is also even. In particular, Abelian spin Hall insulators with protected edge modes must have $1/e^*$ being odd. \label{thm1}
\end{thm}
The second half of the theorem is obtained by combining the first half and the stability criterion in Sec.~\ref{stability}. In the case of $s^z$ conserving Abelian spin Hall insulators, i.e., those with $W=0$, the theorem was previously obtained in Reference~\onlinecite{levin09}.

To prove the theorem, we first simplify the expressions of $e^*$ and $\chi^T\mathcal K^{-1}\tau$. With (\ref{a1}), we find
\begin{align}
e^* &= {\rm min}_{l}[l^T (K-W)^{-1} t], \label{a2}\\
\chi^T\mathcal K^{-1}\tau & = -t^T(K-W)^{-1}t.\label{a3}
\end{align}
Examining these equations, we see that $e^*$ looks like the smallest charge of a quantum Hall system with a $K$-matrix $\mathcal K=K-W$, and $-\chi^T \mathcal K^{-1}\tau$ looks like the Hall conductance. We will find this analogy is useful in our proof, but there is a problem: a $K$-matrix is symmetric while $K-W$ is not. Therefore, we will now discuss how to extend the definitions of charge and statistical phase to a general nonsingular integer matrix $\mathcal K$.  The new charges and statistical phases are mathematically well defined and will help us to complete the proof of the theorem, though their physical meaning is unclear.

Consider a nonsingular integer matrix $\mathcal K$ and a charge vector $t$. Quasiparticles are described by an integer vector $l$. We define the ``left-charge'' of $l$ by
\begin{equation}
q_l=t^T\mathcal K^{-1} l.
\end{equation}
(Similarly, one may define right-charge $\bar q_l=l^T\mathcal K^{-1} t$.) We define mutual statistical phase between quasiparticles $l$ and $l'$ as
\begin{equation}
\theta_{l'l} = 2\pi l'^T\mathcal K^{-1} l.
\end{equation}
Note that if $\mathcal K$ is not symmetric, $\theta_{ll'}\neq \theta_{l'l}$. Physically, the symmetry $\theta_{ll'} = \theta_{l'l}$ must be satisfied. Therefore, the above definition of statistical phase is purely a mathematical construction, without a clear physical meaning.

We will say a quasiparticle $l$ is ``left-trivial'' if the statistical phase $\theta_{l'l}$ is a multiple of $2\pi$ for any quasiparticle $l'$. (Similarly, one may define a right-trivial quasiparticle which has its statistical phase $\theta_{ll'}$ being a multiple of $2\pi$ for any $l'$.) One can show that left-trivial particles are described by vectors $\mathcal K \Lambda$ where $\Lambda$ is an integer vector. They carry integer left-charge $t^T\Lambda$. An important property of a left-trivial particle $l$ is that if $\mathcal K_{II} \equiv t_I \, ({\rm mod}\,2)$, its self-statistical phase $\theta_{ll}$ satisfies
\begin{equation}
\frac{1}{2\pi}\theta_{ll} \equiv q_l \, ({\rm mod}\,2). \label{a6}
\end{equation}
This property is a natural extension of the following property of physical electronic systems: fermionic excitations carry odd charge and bosonic excitations carry even charge.

With the above preparation, we can now prove the theorem. Consider a fictitious ``quantum Hall liquid'' with $\mathcal K = (K-W)$ and a charge vector $t$, following the notations in (\ref{a1}).  We study properties of the particular quasiparticle labeled by the vector $t/e^*$, where $e^*$ is given by (\ref{a2}).  Its properties will lead us to {\bf Theorem \ref{thm1}}. First, this quasiparticle is left-trivial, from the very definition of $e^*$.  Second, its left-charge $Q$ and  self-statistical phase $\theta$ are given by
\begin{align}
Q & = \frac{1}{e^*}t^T(K-W)^{-1}t \label{a8}, \\
\theta & = 2\pi \frac{t^T}{e^*}(K-W)^{-1}\frac{t}{e^*}=2\pi Q \frac{1}{e^*}.
\end{align}
Left-trivial particles all have integer left-charge, so $Q$ is an integer. Then, $\theta/2\pi$ is even if $1/e^*$ is even. Third, with the skew-symmetry of $W$, we have $ (K-W)_{II}=K_{II} \equiv t_I \, ({\rm mod}\,2)$. Then, the property (\ref{a6}) is applicable, implying that $Q$ and $\theta/2\pi$ have the same parity. Thus, $Q$ is even if $1/e^*$ is even. Finally, according to the expression (\ref{a3}), we have $\frac{1}{e^*} \chi^T\mathcal K \tau = - Q$. So, $\frac{1}{e^*} \chi^T\mathcal K \tau$ is even if $1/e^*$ is even. This completes our proof.

In the case of $s^z$ conserving Abelian spin Hall insulators, i.e. $W=0$, the above fictitious quantum Hall liquid with $\mathcal K=K-W$ becomes a real quantum Hall liquid. Then, the above proof has a physical interpretation in terms of a flux insertion thought experiment\cite{levin09}.

A final remark: if we consider general Abelian insulators beyond Abelian spin Hall insulators, the theorem  will break down. A counter example is:
\begin{align}
\cal K & = \left(
\begin{array}{cccc}
0 & 2 & 1 & 1\\
2 &0 & 3 & -3\\
1 & 3 & 1 & 0\\
1& - 3 & 0 & -1
\end{array}
\right),\
\tau = \left(
\begin{array}{c}
0\\
4\\
1\\
1
\end{array}
\right),\nonumber\\
\cal T & = \left(
\begin{array}{cccc}
-1 & 0 & 0 & 0\\
0 & 1 & 0 & 0\\
0 & 0 & 0 & 1\\
0& 0 & 1 & 0
\end{array}
\right),\
\chi = \left(
\begin{array}{c}
0\\
0\\
0\\
1
\end{array}
\right).
\end{align}
It is easy to check that the smallest charge $e^*$ is $1/2$. However, $\chi^T\mathcal K^{-1}\tau/e^*=-1$.


\begin{thebibliography}{100}
\bibitem{kane05a} C. L. Kane and E. J. Mele, Phys. Rev. Lett. {\bf 95}, 226801 (2005).
\bibitem{kane05b} C. L. Kane and E. J. Mele, Phys. Rev. Lett. {\bf 95}, 146802 (2005).
\bibitem{bernevig06} B. A. Bernevig and S.-C. Zhang, Phys. Rev. Lett. {\bf 96}, 106802 (2006).
\bibitem{hasan10} M. Z. Hasan and C. L. Kane, Rev. Mod. Phys. {\bf 82}, 3045 (2010).
\bibitem{xu06} C. Xu and J. E. Moore, Phys. Rev. B {\bf 73}, 045322 (2006).
\bibitem{wu06} C. Wu, B. A. Bernevig, and S. C. Zhang, Phys. Rev. Lett. {\bf 96}, 106401 (2006).
\bibitem{schnyder08} A. P. Schnyder, S. Ryu, A. Furusaki, and A. W. W. Ludwig, Phys. Rev. B {\bf 78}, 195125 (2008).
\bibitem{kitaev09} A. Kitaev, AIP Conf. Proc. {\bf 1134}, 22-30 (2009).
\bibitem{chen13} X. Chen, Z.-C. Gu, Z.-X. Liu, and X.-G. Wen, Phys. Rev. B {\bf 87}, 155114 (2013).
\bibitem{lu12}  Y. -M. Lu and A. Vishwanath, Phys. Rev. B {\bf 86}, 125119 (2012).
\bibitem{levin09} M. Levin and A. Stern, Phys. Rev. Lett {\bf 103}, 196803 (2009).
\bibitem{levin12} M. Levin and A. Stern, Phys. Rev. B {\bf 86}, 115131 (2012).
\bibitem{neupert11}T. Neupert, L. Santos, S. Ryu, C. Chamon, and C. Mudry, Phys. Rev. B {\bf 84}, 165107 (2011).
\bibitem{lu13} Y. -M. Lu and A. Vishwanath, arXiv:1302.2634 (unpublished).
\bibitem{fidkowski10}L. Fidkowski and A. Kitaev, Phys. Rev. B {\bf 81}, 134509 (2010).
\bibitem{ryu12}S. Ryu and S.-C. Zhang, Phys. Rev. B {\bf 85}, 245132 (2012).
\bibitem{yao12}H. Yao and S. Ryu, arXiv:1202.5805.
\bibitem{qi13}X.-L. Qi, New J. Phys. {\bf 15}, 065002 (2013).
\bibitem{gu13}Z.-C. Gu and M. Levin, arXiv:1304.4569.
\bibitem{vishwanath13}A. Vishwanath and T. Senthil, Phys. Rev. X {\bf 3}, 011016 (2013).
\bibitem{footnote1}The same terminology ``weak symmetry breaking'' was used by A. Kitaev to describe nontrivial behavior of anyonic excitations under symmetry transformation in the appendix F of Ann. Phys. {\bf 321}, 2 (2006). It is not clear if there is a connection between that notion of weak symmetry breaking and the phenomena studied in this paper.
\bibitem{wen} X.-G. Wen, {\it Quantum Field Theory of Many-Body Systems} (Oxford, 2004).
\bibitem{wen92} X.-G. Wen and A. Zee, Phys. Rev. B {\bf 46}, 2290 (1992).
\bibitem{wen95} X.-G. Wen, Adv. Phys. {\bf 44}, 405 (1995).
\bibitem{haldane95} F. D. M. Haldane, Phys. Rev. Lett. {\bf 74}, 2090 (1995).
\bibitem{jwang13}J. Wang and X.-G. Wen, arXiv:1212.4863.
\bibitem{wen89}X.-G. Wen, Phys. Rev. B {\bf 40}, 7387 (1989).
\bibitem{levin13} M. Levin, Phys. Rev. X {\bf 3}, 021009 (2013).
\bibitem{fu08} L. Fu and C.L. Kane, Phys. Rev. Lett. {\bf 100}, 096407 (2008).
\bibitem{metlitski13}M. A. Metlitski, C. L. Kane, and M. P. A. Fisher, arXiv:1306.3286.
\bibitem{wang13} C. Wang, A. C. Potter, and T. Senthil, arXiv:1306.3238.
\bibitem{bonderson13} P. Bonderson, C. Nayak, and X.-L. Qi, arXiv:1306.3230.
\bibitem{chen13b} X. Chen, L. Fidkowski, and A. Vishwanath, arXiv:1306.3250.
\bibitem{lu-lee}  Y.-M. Lu and D.-H. Lee,  arXiv:1311.1807.


\end{thebibliography}
\end{document}